\documentclass[letterpaper,twocolumn,10pt]{article}
\usepackage{usenix,epsfig,endnotes}
\usepackage{endnotes}
\usepackage{times}
\usepackage{balance}
\usepackage{amsfonts}
\usepackage{epsfig}
\usepackage{amssymb}
\usepackage{amscd}
\usepackage{amsmath}
\usepackage{microtype}
\usepackage{array}
\usepackage{graphics}
\usepackage{epsfig}
\usepackage{multirow}
\usepackage{multicol}
\usepackage{makecell}

\usepackage{caption}
\usepackage{listings}
\usepackage{graphicx}
\usepackage{subcaption}
\usepackage{amsmath}
\usepackage{amsfonts}

\usepackage{listings}
\usepackage[hyphens]{url}
\usepackage{soul}

\usepackage{wrapfig}

\usepackage{enumitem}

\usepackage{diagbox}

\usepackage{tabularx}
\usepackage{epstopdf}
\usepackage{threeparttable}
\usepackage{setspace}
\usepackage{lipsum}
\usepackage{color,xcolor}
\usepackage{color, colortbl}
\definecolor{lgray}{gray}{0.95}

\usepackage{cite}

\usepackage{pifont}
\usepackage[labelfont=bf, font=footnotesize]{caption}
\newcommand{\ignore}[1]{}
\newcommand{\revised}[1]{}

\newcommand{\pr}[1]{\mathrm{Pr}\{#1\}} 
\newcommand{\thresh}{\mathrm{t}}

\newcolumntype{M}[1]{>{\centering\arraybackslash}m{#1}}
\newcolumntype{C}[1]{>{\centering\let\newline\\\arraybackslash\hspace{0pt}}m{#1}}
\newcolumntype{L}[1]{>{\raggedright\let\newline\\\arraybackslash\hspace{0pt}}m{#1}}

\newcommand{\specialcell}[2][c]{%
  \begin{tabular}[#1]{@{}c@{}}#2\end{tabular}}
\usepackage{lipsum}
\usepackage{titlesec}

\titlespacing\section{0pt}{6pt plus 2pt minus 1pt}{0pt plus 2pt minus 2pt}
\titlespacing\subsection{0pt}{6pt plus 2pt minus 1pt}{0pt plus 2pt minus 2pt}
\titlespacing\subsubsection{0pt}{0pt plus 2pt minus 2pt}{0pt plus 2pt minus 2pt}

\usepackage{enumitem}
\setlist{nosep}

\usepackage{float}
\usepackage{authblk}

\usepackage{algorithm}
\usepackage[noend]{algpseudocode}

\date{}

\title{Understanding Membership Inferences on \\ Well-Generalized Learning Models}

\author[1]{Yunhui Long}
\author[1]{Vincent Bindschaedler}
\author[2]{Lei Wang}
\author[2]{Diyue Bu}
\author[2]{Xiaofeng Wang}
\author[2]{Haixu Tang}
\author[1]{Carl A. Gunter}
\author[3,4]{Kai Chen}
\affil[1]{University of Illinois at Urbana-Champaign}
\affil[2]{Indiana University Bloomington}
\affil[3]{State Key Laboratory of Information Security, Institute of Information Engineering, Chinese Academy of Sciences}
\affil[4]{School of Cyber Security, University of Chinese Academy of Sciences}

\begin{document}
\maketitle

\subsection*{Abstract}
Membership Inference Attack (MIA) determines the presence of a record in a machine learning model's training data by querying the model. Prior work has shown that the attack is feasible when the model is overfitted to its training data or when the adversary controls the training algorithm. However, when the model is \emph{not} overfitted and the adversary does \emph{not} control the training algorithm, the threat is not well understood. In this paper, we report a study that discovers overfitting to be a \textit{sufficient} but not a \textit{necessary} condition for an MIA to succeed. More specifically, we demonstrate that even a well-generalized model contains vulnerable instances subject to a new generalized MIA (GMIA). In GMIA, we use novel techniques for selecting vulnerable instances and detecting their subtle influences ignored by overfitting metrics. Specifically, we successfully identify individual records with high precision in real-world datasets by querying black-box machine learning models. 
Further we show that a vulnerable record can even be \textit{indirectly} attacked by querying other related records and existing generalization techniques are found to be less effective in protecting the vulnerable instances. Our findings sharpen the understanding of the fundamental cause of the problem: the unique influences the training instance may have on the model. 

\section{Introduction}

The recent progress on machine learning has brought in a new wave of technological innovations, ranging from automatic driving, face recognition, natural language processing to intelligent marketing, advertising, healthcare data management, etc. To support the emerging machine learning ecosystem, major cloud providers are pushing \textit{Machine Learning as a Service} (MLaaS), providing computing platforms and learning frameworks to help their customers conveniently train their own models based upon the datasets they upload. Prominent examples include Amazon Machine Learning (ML), Google Prediction API, and Microsoft Azure Machine Learning. The models trained on these platforms can be made available by the data owners to their users for online queries. What is less clear are the privacy implications of these exported models, particularly whether uncontrolled queries on them could lead to exposure of training data that often includes sensitive content such as purchase preferences, patients' health information, and recorded commands and online behavior.    

\vspace{2pt}\noindent\textbf{Membership inference attack}. Prior research demonstrates that a \textit{membership inference attack} (MIA) can succeed on \emph{overfitted} models with only black-box access to the model~\cite{shokri2016membership}. In such an attack, the adversary, who can only query a given \textit{target model} without knowing its internal parameters, can determine whether a specific record is inside the model's training dataset. This type of attacks can have a significant privacy implication such as re-identifying a cancer patient whose data is used to train a classification model. For this purpose, the prior research trains an attack model that utilizes the target model's classification result for a given input to determine whether the input is present in the target model's training set. Such an attack model can be constructed using labeled datasets generated by a set of \textit{shadow models} trained to imitate the behaviors of the target model. This approach is effective when the target models are \textit{overfitted} to training data.


It remains unclear whether it is feasible to perform MIA on \textit{well-generalized} models with \emph{only} black-box access. This problem should be distinguished from prior attacks on non-overfitted models under \emph{a different adversarial model}. In these attacks, the adversary controls the training algorithm and stealthily embeds information in the model~\cite{song2017machine,yeom2017unintended}.

\vspace{2pt}\noindent\textbf{Rethinking ML privacy risk}. In our research, we revisited the threat of MIA, in an attempt to answer the following questions: (1) is overfitting a root cause of membership disclosure from a machine learning model?  (2) if so, is generalization the right answer to the problem?  (3) if not, what indeed causes the information leak? The findings of our study, though still not fully addressing these issues, make a step closer toward that end, helping us better understand the threat of MIA.  

\noindent\textit{(1) Is overfitting a root cause of membership disclosure from a machine learning model?} 

We discover that overfitting, as considered in evaluating machine learning models, can be \textit{sufficient} but is by no means \textit{necessary} for exposing membership information from training data. As evidence, we run a new MIA (called generalized MIA or GMIA) that successfully identifies \textit{some} individuals in the training sets from three neural-network models, for predicting salary class, cancer diagnosis and written digits, \textit{even when these models are not overfitted}. 
Particularly, our attack automatically picks $5$ vulnerable patient samples from the Cancer dataset~\cite{Lichman:2013}, $16$ images from the MNIST dataset~\cite{lecun2010mnist}, and $13$ individuals from the Adult dataset~\cite{Lichman:2013} as the attack object. 
We identified $73.88\%$ of the models, from which the target images in the MNIST dataset can be inferred with a precision of $93.36\%$.
Similarly, we inferred the presence of target patients in the Cancer dataset with a precision of $88.89\%$ in $3.2\%$ of the models and the presence of target individuals in the Adult dataset with a precision of $73.91\%$ in $5.23\%$ of the models. 

Further interesting is the observation that the adversary does not even need to query the models for the target record to determine its presence, as it does in the prior research: instead, the adversary can search for different but related records and use their classifications by models to determine the object's membership in the training data. 

\noindent\textit{(2) Is generalization the right solution for membership disclosure?}

We find that existing regularization approaches are \textit{insufficient} to defeat our attack, which can still determine the presence of an image in the MNIST dataset in $34\%$ of all the models with a precision of $100\%$ even when L2 regularization is applied. This finding deviates from what is reported in the prior research, whose MIA can be effectively suppressed by regularization~\cite{shokri2016membership}. 

\noindent\textit{(3) What is the fundamental cause of membership disclosure?}

We observe that such information leaks are caused by the unique influences a specific instance in the training set can have on the learning model. The influences affect the model's outputs (i.e., predictions) with regards to a single or multiple inputs. So once the adversary has asked enough questions, no guarantee will be there that he cannot capture the influences (possibly from multiple queries) to infer the presence of a certain training instance.
It is important to note that overfitting is essentially a \textit{special} case of such unique influences but the generic situation is much more complicated. In detection of overfitting, we look for an instance's positive impact on a model's accuracy for the training set and limited/negative impact for the testing set. On the contrary, finding the unique influences in general needs to consider the case when an instance both contributes useful information to the model for predicting other instances and brings in noise uniquely characterizing itself. The model generalization methods that suppress overfitting may reduce the noise introduced by training instances, but cannot completely remove their unique influences, particularly the influences essential for the model's prediction power. On the other hand, noise adding techniques based on the concept of differential privacy~\cite{dwork2006differential} can guarantee the low influence of each training instance, while also reduces the prediction accuracy of the model. How to capture the non-noise influences of learning instances through the model's output and how to identify all the {\em vulnerable} instances with identifiable influences on the model remain open questions. 


\vspace{2pt}\noindent\textbf{Generalized MIA}. 
These discoveries are made possible by a novel inference attack we call \textit{Generalized MIA (GMIA)}. In GMIA, we propose a new technique for identifying vulnerable records in a large dataset and new methods for detecting the small influence of these records that are ignored by overfitting metrics and the prior attack. 

Unlike overfitted models, whose answers (probabilities) to the queries on the training instances differ significantly from those to other queries, a well-generalized model behaves similarly on the training data and test data. As a result, no longer can we utilize shadow models to generate a meaningful training set for the attack model, since most positive instances here (those inside shadow models' training sets) can be less distinguishable from the negative instances (not in their training sets), in terms of their classification probabilities. To address this challenge, our approach focuses on detecting and analyzing vulnerable target records (outliers) to infer their membership. More specifically, GMIA first estimates whether a given instance is an outlier with regards to the data accessible to the adversary. This estimation is done by extracting high-level feature vector from the intermediate outputs of models trained on these data. We believe that an outlier is more likely to be a vulnerable target record when it is indeed inside the target model's training set. Then we train a set of \textit{reference} models without the target record in the training set, and use these models to build the distribution for the target record's classification probabilities. After that, we run a hypothesis test to decide whether its classification by the target model is in line with this distribution. This approach successfully identifies training records of well-generalized models. For example, on the MNIST dataset, our attack achieves a precision of $93.36\%$ in $73.88\%$ of the models (on the vulnerable objects) when the cutoff $p$-value is $0.01$.

It is even more challenging to attack a target record without directly querying it (\emph{an indirect inference}), which has \textit{never} been done before. To find the right queries for the object, GMIA trains two sets of reference models: those include the object (\emph{positive reference models}) and those do not (\emph{reference models}). These two sets of models are used to filter out random queries, finding those whose probabilities for receiving the object's class labels are almost always higher with the positive reference models than with the reference models. The selected queries are run against the target model, and their results (classification probabilities) are compared with their individual distributions built from the \textit{reference models} through a set of hypothesis tests. Finally, the test results ($p$-value) of individual queries are combined using Kost's method~\cite{kost2002combining} to  determine the object's presence in the target model's training set. On the Adult dataset, with a cut-off $p$-value of $0.01$, this indirect attack inferred the presence of a record with a precision of $100\%$ in $16\%$ of the models when a direct attack failed to infer any of them. 

\vspace{2pt}\noindent\textbf{Contributions}. The contributions of the paper are summarized as follows:

\vspace{2pt}\noindent$\bullet$\textit{ New understanding about generalization and privacy}. We revisit the membership inference problem and find that overfitting is not a necessary condition for information leaks: even a well-generalized model still cannot prevent MIA, whenever some of its training instances have unique impacts on the learning model. This discovery reveals the fundamental challenges in protecting data privacy for machine learning models and can potentially inspire follow-up research on mitigating this risk. 

\vspace{2pt}\noindent$\bullet$\textit{ New techniques for membership inference attacks}. We present new techniques for membership inferences on a well-generalized model. Our approach addresses the challenges of finding vulnerable target records, identifying their small influences on the target model, and attacking the target model without directly querying the target records. 

\vspace{2pt}\noindent$\bullet$\textit{ Implementation and evaluation}.  We implement our attacks and evaluate them against real-world datasets. Our studies demonstrate their effectiveness and also highlight the challenges in protecting machine learning models against such threats. 

%
\begin{figure*}[!t]
    \centering
    \vspace{-30pt}
         \begin{subfigure}[b]{0.305\textwidth}
    \includegraphics[width=\textwidth]{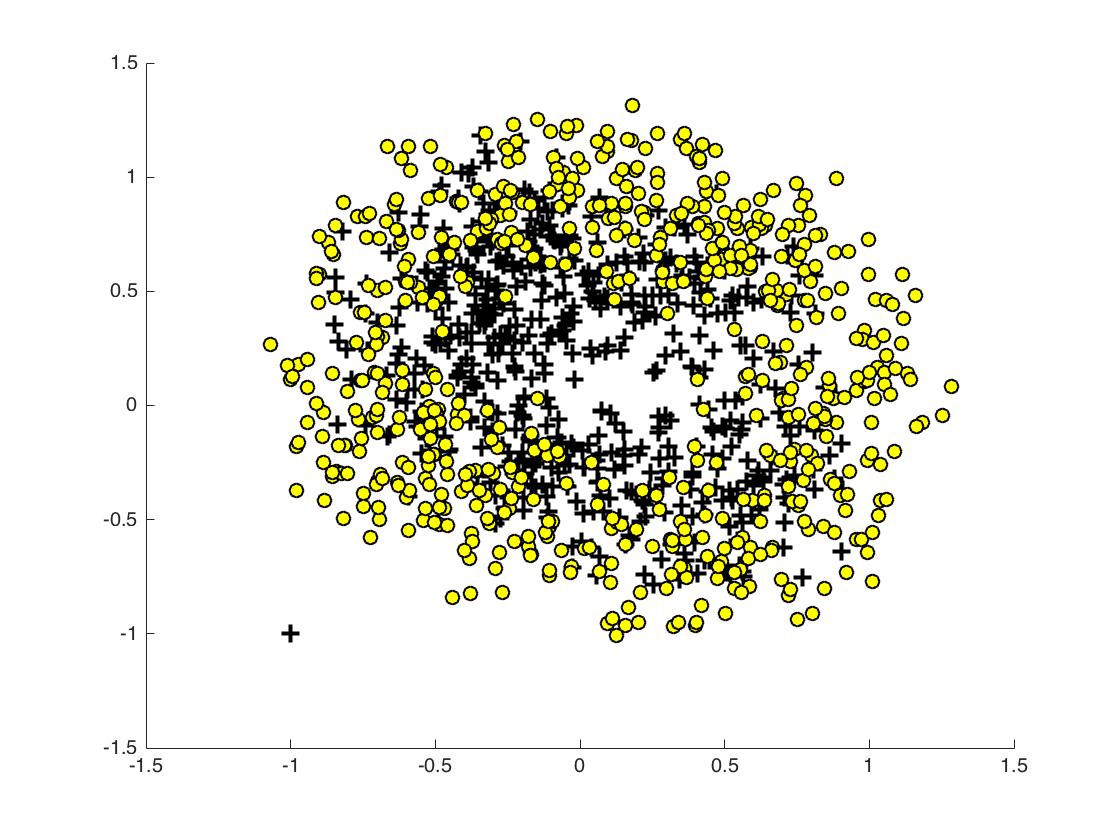}
\caption{The toy example dataset. The dataset is composed records with real-valued features (x-axis and y-axis) and a binary label ($+$ or $-$) used for classification. 
\label{fig:uniqinf:toydataex}}  
    \end{subfigure}\hfill    
     \begin{subfigure}[b]{0.305\textwidth}
    \includegraphics[width=\textwidth]{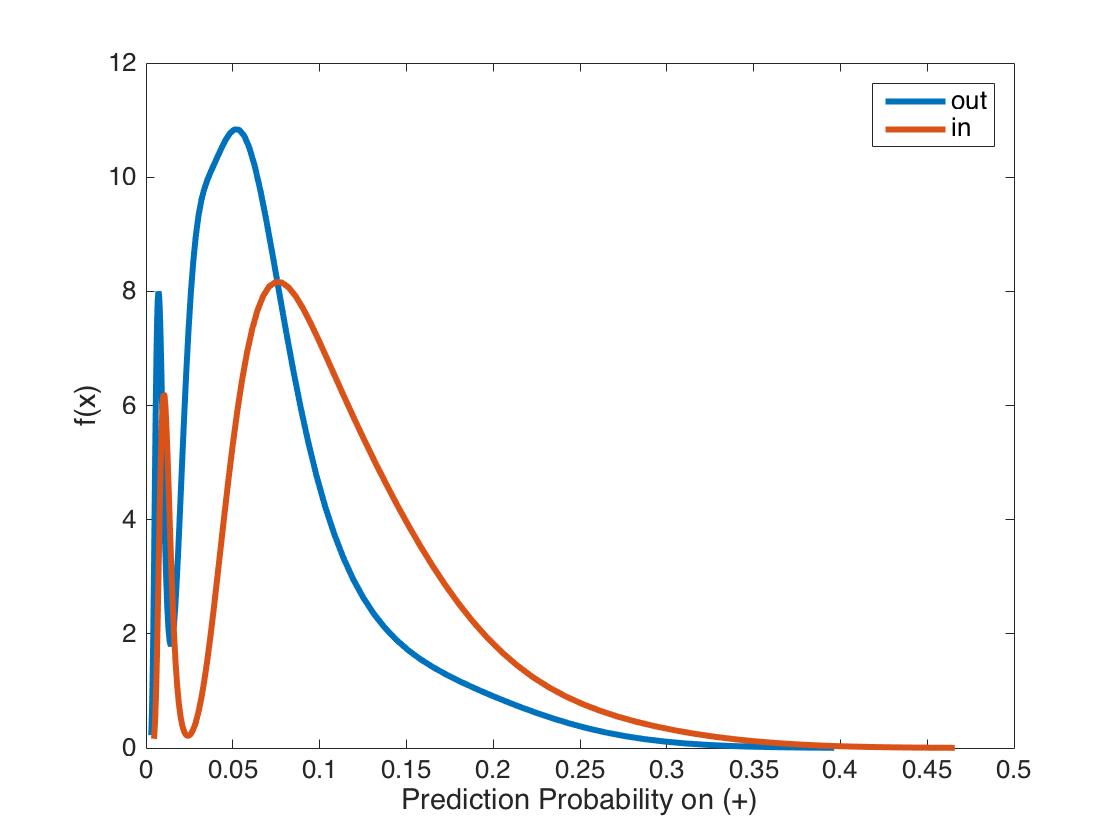}
\caption{\small Influence on the probability density function (${\rm f}(x)$) of the model's prediction. The \texttt{in} and \texttt{out} distributions do not fully overlap, which allows the adversary to distinguish them.
\label{fig:uniqinf:ex}}  
    \end{subfigure}\hfill    
    \begin{subfigure}[b]{0.305\textwidth}
\includegraphics[width=\textwidth]{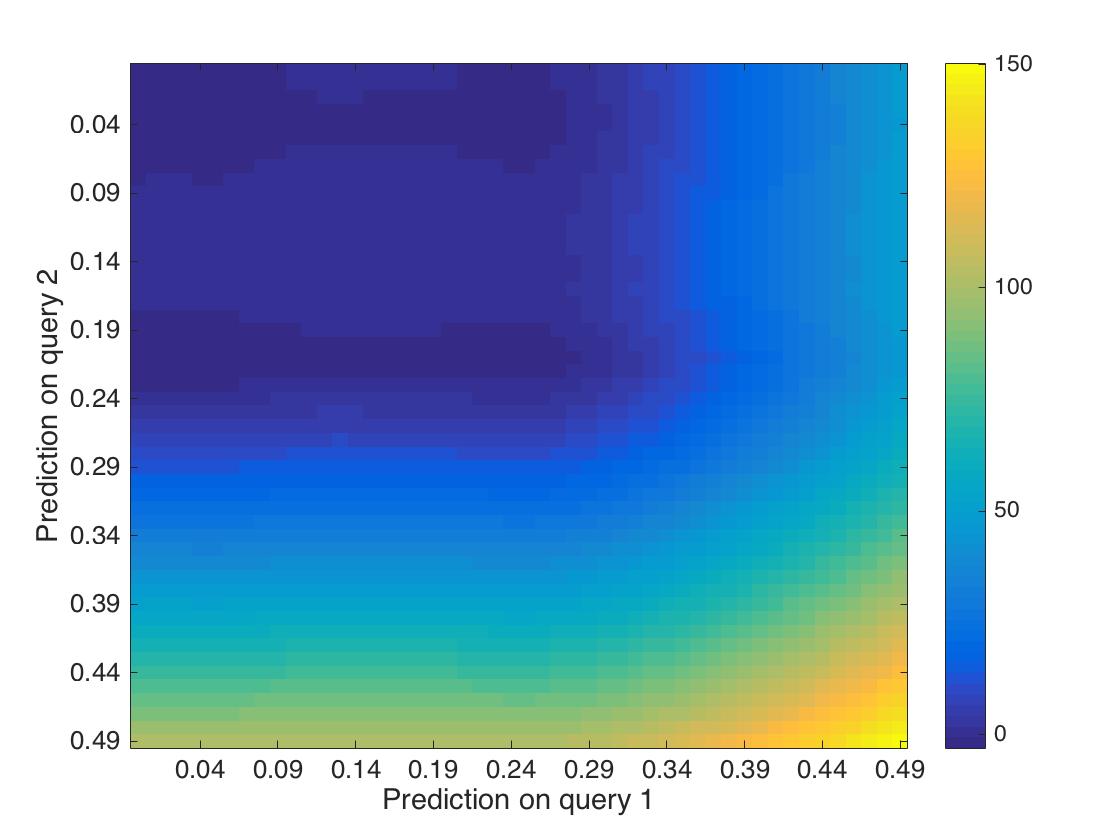}
\caption{Estimation of $\frac{\Pr_{\rm in}}{\Pr_{\rm out}}$ based on the model's predictions for two queries (query 1 on x-axis and query 2 on y-axis). 
When prediction for both queries are above $0.4$, it is significantly more likely than not that the record was part of the training data.
\label{fig:uniqinf:indirect}}
    \end{subfigure}
    \vspace{8pt}
    \caption{Understanding the unique influence of a record through a toy example dataset (a). The adversary performs MIA by fingerprinting the target record's influence on the model's outputs (predicted class probabilities). There are two competing hypotheses: (1) $H_{\rm in}$: record is part of the training data, and (2) $H_{\rm out}$: record is not part of the training data. The adversary infers membership status by estimating which hypothesis is more likely based on the model's outputs.
    }
\vspace{-5pt}
\end{figure*}

\section{Background}
%

\subsection{Membership Inference Attacks}
In a membership inference attack, the adversary's goal is to infer the membership status of a target individual’s data in the input dataset to some computation. For a survey, the adversary wishes to ascertain, from aggregate survey responses, whether the individual participated in the survey. For machine learning, the adversary wishes to ascertain whether the target's record was part of the dataset used to train a specific model. 

One of the first prominent examples of membership inference attacks occur in the context of Genome-Wide Association Studies (GWAS). The seminal work of Homer et al.~\cite{homer2008resolving} show that $p$-values, a type of aggregated statistics routinely published when reporting the results of studies, could be used to successfully infer membership status. Although this attack requires that the adversary know the genome of the target individual, it teaches an important lesson: seemly harmless aggregate statistics may contain sufficient information for successful membership inferences. As a consequence of this attack, NIH has removed all aggregate data of GWAS from public websites~\cite{zerhouni2008protecting}.

More recently, it was shown that membership inference attacks can occur in the context of machine learning. Shokri et al.~\cite{shokri2016membership} demonstrated that an adversary with only black-box access to a classifier could successfully infer membership status. However, their attack only works when the classifier is highly overfitted to its  training dataset. 

\subsection{Model Generalization}
A desirable property of any model is having low generalization error. Informally, this means the model should have good performance on unseen examples. More precisely, we adopt the approach of~\cite{bousquet2002stability} which defines generalization error as the expected model error with the expectation taken over a random example from the population. 

In practice, one does not have access to the population but only to samples from it. Thus, we must estimate the generalization error empirically. To do this, we can simply measure the generalization error with respect to a hold out set instead of the population. Informally, a good indication of low generalization error is if the model's performance is (almost) the same on the training and testing datasets.

\ignore{
\subsection{Differential Privacy}
Differential privacy~\cite{dwork2006differential} is a prominent way to formalize privacy against membership inference. An algorithm $F$ satisfies $\varepsilon$-{\em differential privacy} if for any neighboring datasets $D$, $D'$, and any output $S \subseteq {\rm{Range}}(F)$:
	\[ \pr{F(D') \in S} \leq e^\varepsilon \pr{F(D) \in S} \ . \]
%
%
In the case of machine learning models, the output distribution is over all possible models, and it is the process of training the model training that must satisfy differential privacy. Informally, if adding or removing a record only has a (small) bounded influence on what an adversary observes, then no membership inference attack can succeed by a probability more than the bound allows.

Notably, there are no generic methods to achieve differential privacy for all useful machine learning models. More importantly, even if these methods are developed, their applications to real-world machine learning problems may significantly decrease the accuracy of the models, and thus will reduce their utility~\cite{alvim2011differential}. As a result, machine learning practitioners seldom use differential privacy.}

\subsection{Adversary Model}
We consider an adversary mounting a membership inference attack against an already trained machine learning model, in which the adversary attempts to infer if a target record $r$ is used as a training record for the target model $M$. 
We assume that the adversary has black-box access to the target model, i.e., he can issue arbitrary queries and retrieve the answers (e.g., the probability vector) from the model; the number of queries, however, may be limited.
Similar as the previous work~\cite{shokri2016membership}, we further assume that the adversary either (1) knows the structure of the target model (e.g., the depth and the number of neurons each layer of the neural network) and the training algorithm used to build the model, or (2) has black-box access to the machine learning algorithm used to train the model. We also assume that the adversary has some prior knowledge about the population from which the training records are drawn. Specifically, the adversary can access a set of records that are drawn independently from that population, which may or may not overlap with the actual training data for the target models; but the adversary does not have any additional information about whether these records are present in the training data. These records can often be obtained from public dataset with similar attributes or from previous data breaches. 
\section{Understanding Membership Inference Attacks\label{sec:understanding}}
Previous work demonstrates the vulnerability of machine learning (ML) models to membership inference attacks (MIAs), but little was known about its root cause. We revisit this based on the new results presented in this paper, in an attempt to understand the source of information leaks in machine learning models that can be exploited by MIAs.

\subsection{Overfitting and Vulnerability of Machine Learning Models}
%
Shokri et al.~\cite{shokri2016membership} show that overfitting is a sufficient condition for MIA. But is overfitting necessary for membership inference? This question is crucial yet not answered by prior work. If overfitting is a prerequisite for successful MIA then attacks can be mitigated using techniques to enforce generalization (e.g., model regularization). 

Our research leads to the new observations that MIAs can still succeed even when the target model is well generalized. 
For example, using the MNIST set, we find 16 records (out of 20,000) whose membership can be inferred successfully with greater than $90\%$ precision in $74\%$ of the models. Moreover, we find that, while model regularization improves the generalization, it does not reliably eliminate the threat. For instance, using an image dataset, after applying L2 regularization with a coefficient of $0.01$, the membership status of one image can still be inferred with $100\%$ precision in $34\%$ of the models. 

\subsection{Influence and Uniqueness}
What causes models to leak membership information? Our research uncovers a new way of thinking about this question in terms of the unique influence of vulnerable records on the model. Informally, a record has a unique influence if there exists a set of queries for which the model outputs can reliably reveal the record's presence in the training data. In such a case, the adversary effectively infers the membership of a target record by the {\em fingerprint} of the record, i.e., the model outputs to queries of the target record and relevant records when the target record is included in the training set of the target model. 

To explain why the unique influence of a target record is the key for a successful MIA, we consider an adversary attempting to determine the membership status of a target record $r$ through black-box access to a target model $M$, using hypothesis testing between two hypotheses:
\begin{itemize}
[labelindent=12pt,labelwidth=1cm,itemindent=1.5cm,leftmargin=!,align=left]
	\item[($H_{\rm in}$)]{$r$ is in the training set of $M$}
	\item[($H_{\rm out}$)]{$r$ is {\em not} in the training set of $M$}
\end{itemize}
By querying the model $M$, the adversary gathers evidence in favor of either $H_{\rm in}$ or $H_{\rm out}$, eventually deciding in the favor of the more likely hypothesis. 

To illustrate this approach, we use a toy dataset with 1,181 records (as shown in Figure~\ref{fig:uniqinf:toydataex}) to train a neural network model with two fully connected layers for binary classification. Suppose we want to infer the membership of a record $r$ by querying a record $q$.
Let $M(q)$ be the model’s output to $q$. Over the record space from which the training records are sampled, we derive two probability distributions of the output of $q$ on the two different sets of models, respectively: 1) the models trained with $r$, and 2) the models trained without $r$. Specifically, as shown in Figure~\ref{fig:uniqinf:ex}, the probability density functions (pdfs) of the model outputs (i.e., the output probability of the positive class) under the hypotheses $H_{\rm in}$ and $H_{\rm out}$, respectively,  do not fully overlap, indicating an adversary can decide in favor of $H_{\rm in}$ if the output probability of the positive class is above a threshold (e.g.,  $0.15$).

The key to this strategy is that the two distributions are distinguishable. This happens because for a significant number of models, the outputs on $q$ are consistently different when $r$ is or is not included in the training data. In other words, $r$ has a unique influence on models with respect to the record $q$. In practice, an adversary can only approximate these two distributions; but, as we show in this paper, it is feasible to identify vulnerable records, i.e., those with unique influences on the models. For example, for the MNIST dataset, we can efficiently infer the presence of 16 vulnerable images in the training dataset with precisions greater than $90\%$ in $73.88\%$ of the time.

We stress that what matters here is not the strength of the influence of records, but the influence being unique with respect to other records in the training set and in the record space. The attack fails if there are other records in the record space that would show a similar influence on the model as $r$ if some of them were included in the training set. In such a case, $r$ does not have a {\em unique} influence to the model, and the two distributions largely overlap.

\subsection{Types of Influences}
Our work here also demonstrates distinguishing between different kinds of influences that a record may have on the model, which lead to different mechanisms of MIA. Intuitively, we expect that strongest influence of a record is on the model's output for the query of this specific record. In fact, this is precisely consistent with our experimental observations
However, inclusion of a target record in the training set may influence the model's output behavior on the queries of other records, which may or may not be strongly correlated with the target record with respect to their features. Surprisingly, in our experiments, we observe that the attacks leveraging these {\em indirect} influences (i.e., the {\em indirect inferences}) are sometimes more effective than those based on the direct influences the target records. Specifically, in the Adult dataset, we identify a record whose presence can be inferred by the indirect inference with $100\%$ precision in $14\%$ of the models, whereas the direct inference failed to infer the record's precense in any of the models.

The indirect inferences are powerful because they allow an adversary to accumulate evidence from multiple queries. The more queries the adversary submits, the more likely an adversary can gather the unique influences of the target record, and thus the easier it is to discern the two distributions under different hypotheses as described above. Figure~\ref{fig:uniqinf:indirect} illustrates this concept through a heatmap of the likelihood ratio under the hypotheses of $H_{\rm in}$ and $H_{\rm out}$. 

\section{Generalized Membership Inference Attack}
\label{sec:attack}
%
In this section, we present the major components of the generalized MIA (GMIA) framework: reference model generation, vulnerable records selection, and the inference attacks. The latter includes the \emph{direct inference} which queries the target record and the \emph{indirect inference} which queries selected non-target records. 

\subsection{Attack Overview}
\begin{figure}[t]
\centering
\vspace{-10pt}
\includegraphics[width=0.5\textwidth]{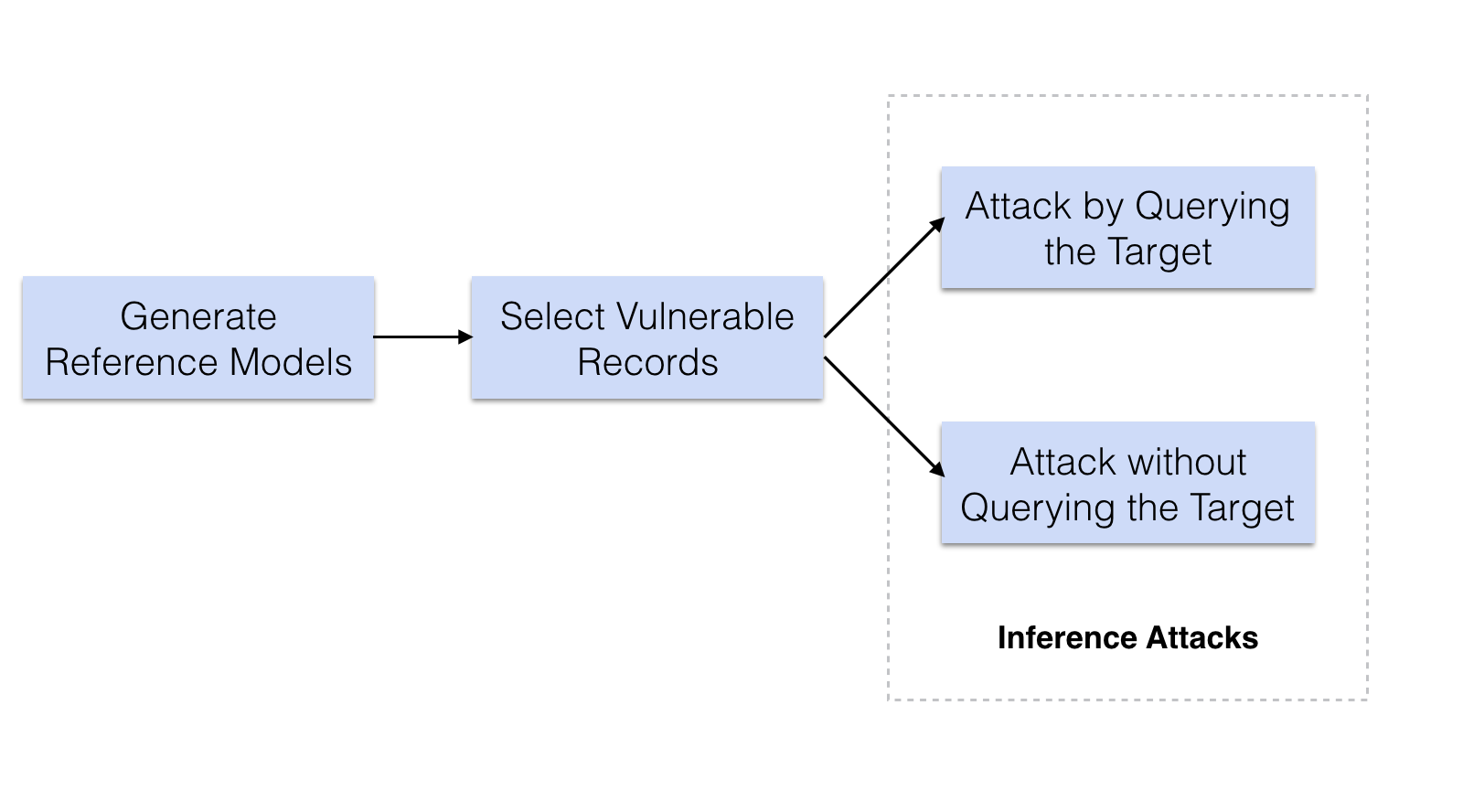}
\vspace{-30pt}
\caption{Attack Overview 
\label{fig:overview}}
\end{figure}

Figure~\ref{fig:overview} shows the attack components in their logical sequence. Below, we briefly describe the methods involved in each component and the motivation. We present the details in the follow-up sections.

\vspace{2pt}\noindent\textbf{Building Reference Models}. %
We build reference machine learning (ML) models to imitate the prediction behaviors of the target model, using reference records accessible to the adversary that represent the space where the actual training data are sampled. As the number of available reference records may be limited, we adopt bootstrap sampling~\cite{efron1982jackknife} to construct training dataset for building multiple reference models. Once constructed, the reference models are exploited in each steps of the GMIA framework, including target record selection, query selection, and hypothesis testing. 

\vspace{2pt}\noindent\textbf{Selecting Vulnerable Target Records}. %
In well-generalized models, not all training records are vulnerable to MIA. Therefore, identifying vulnerable target records is the key to an effective attack. 
We develop a method for selecting vulnerable records by estimating the number of neighbors they have in the the sample space represented by the reference dataset. Records with fewer neighbors are more vulnerable under MIA because they are more likely to impose unique influence on the machine learning models. In order to identify neighbors of a given record, we construct a new feature vector for each record based on the intermediate outputs of reference models on this record, which implies this record's influence on the target machine learning model.

\vspace{2pt}\noindent\textbf{Direct Inference by Querying the Target Record}. %
A training record usually influences the model's predictions on itself. However, in well-generalized models, this influence is usually small and hard to detect. In a direct inference, we attack a machine learning model by submitting a query of the target record. We use a hypothesis test to determine whether the target model's prediction is deviated from the predictions of reference models for that the target records are not used in the training. The $p$-value from the hypothesis test indicates the confidence of the attack and thus allows the adversary to efficiently estimate the performance of the attack.

\vspace{2pt}\noindent\textbf{Indirect Inference without Querying the Target Record }. %
We observe that a training record influences a model's predictions not only on itself but also on other seemingly uncorrelated records (called \emph{enhancing records} in GMIA). In GMIA, we use novel techniques that iteratively search for and select enhancing records. Our indirect inferences using the enhancing records can successfully infer the presence of a target record without querying it. Moreover, the indirect inferences sometimes outperform direct inferences by accumulating more information from multiple queries. 

\subsection{Building Reference Models}
\begin{figure}[t]
\centering
\vspace{-20pt}
\includegraphics[width=0.5\textwidth]{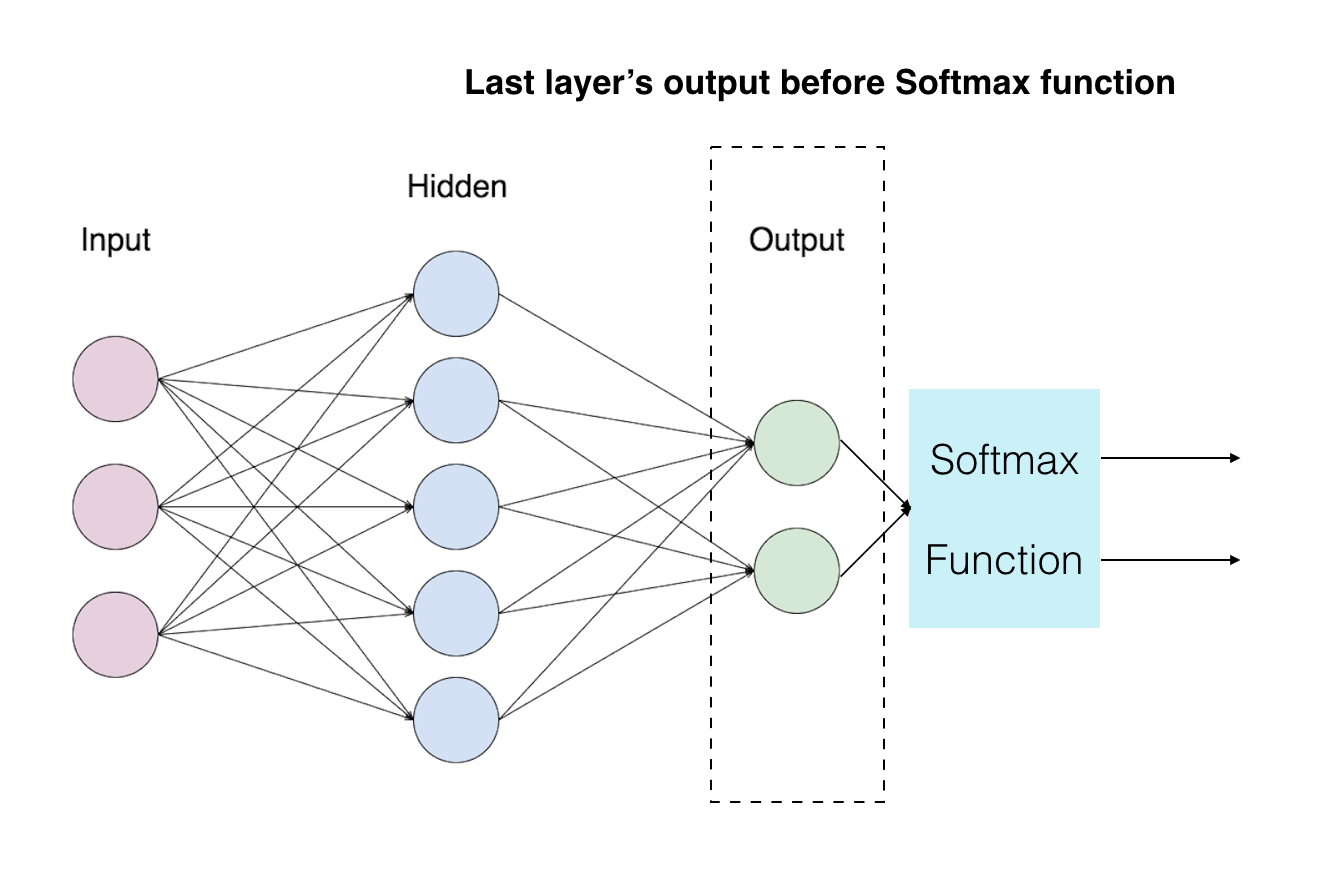}
\vspace{-10pt}
\caption{Last layer output of a two-layer neural network 
\label{fig:dnn}}
\end{figure}

GMIA exploits a target record's unique influence on the outputs of a machine learning model to infer the presence of the record in the training set of the target model (called target training set). To identify such influence, we need to estimate the model's behavior when the target record is \emph{not} in the target training set. To achieve this goal, we build \emph{reference models}, which are trained using the same algorithm on {\em reference datasets} sampled from the same space as the target training set, but not containing the target record. The process of building reference models are illustrated below.

To start with, we need to construct $k$ reference datasets with the same size as the target training set. Since most practical machine learning models are trained on large training datasets, it is difficult for an adversary to get access to an even larger dataset with $k$ times records as the target training set. Consequently, if we build the reference datasets by sampling without replacement from the whole set of reference records, the resulting datasets may share many records, and the reference models built from them would be alike and give similar outputs. To address this issue, we use bootstrap sampling~\cite{efron1982jackknife} to generate the reference datasets, where each dataset is sampled with replacement. Bootstrap sampling reduces overlaps among reference datasets, providing a better approximation of datasets sampled from distribution of the target training set. Each reference dataset is then used to train a reference model using the same training algorithms as used for training the target model. 

\subsection{Selecting Vulnerable Records}
\begin{figure}[t]
\centering
\vspace{-20pt}
\includegraphics[width=0.5\textwidth]{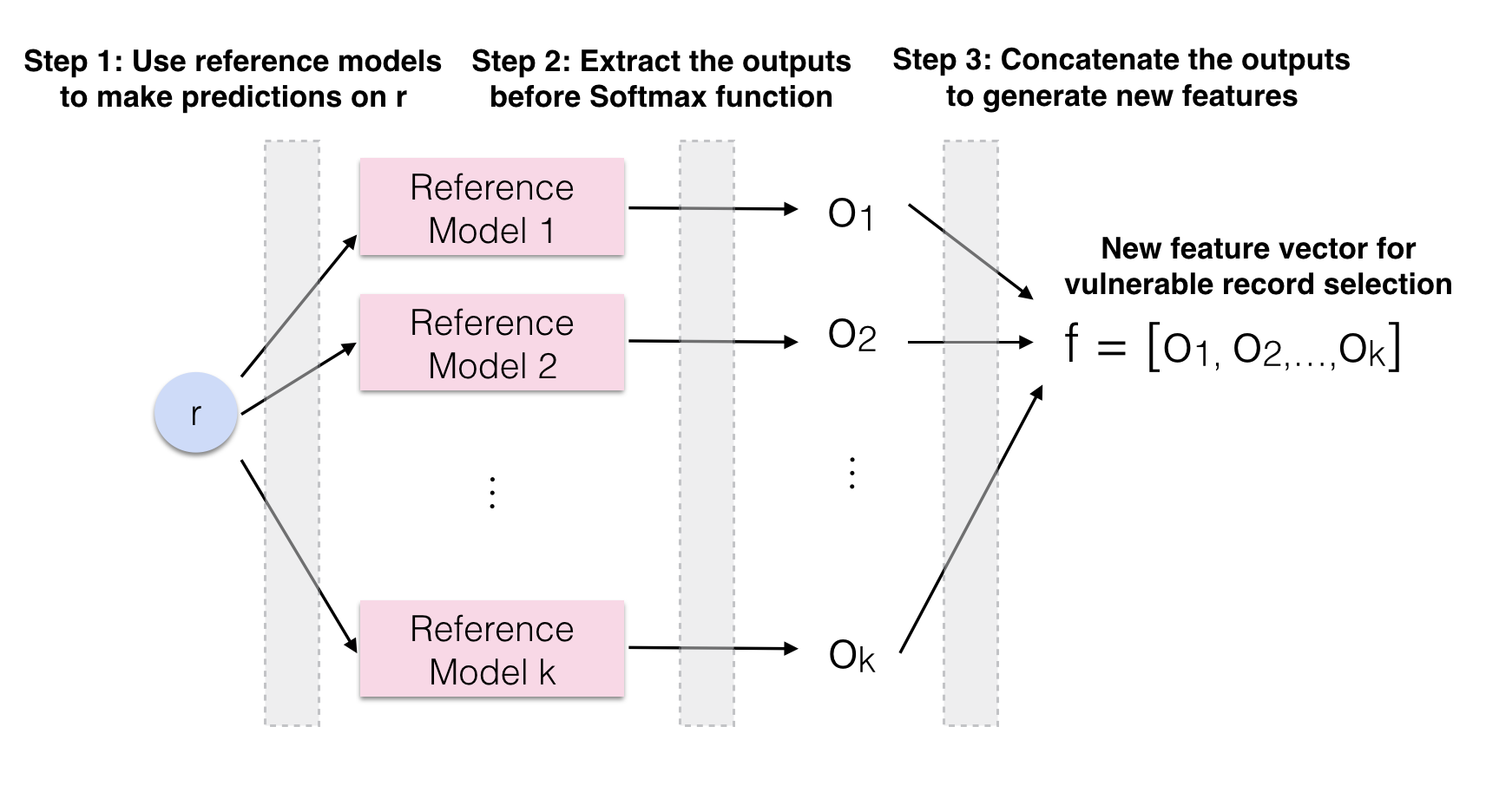}
\vspace{-10pt}
\caption{Generate new features for vulnerable record selection
\label{fig:features}}
\end{figure}

Not all training records are vulnerable to MIA. In an extreme case, if two records are nearly identical, it is difficult to discern which one of them is indeed present in the training dataset because their influence on the model is indistinguishable. In general, we want to measure the potential influence of a target record so as to select vulnerable records with the greatest influences and subject them to MIA in the subsequent steps. 
It is worth noting that, although the the training records imposing unique influence on the model are often {\em outlier records} (i.e., with distinct feature vectors) in the training set, the outlier records do not always have unique influence on the model because the training algorithm may decide that some features should be given higher weights than others and some features should be combined in the model. For example, a neural network trained on hand written digit datasets learns the contour of written digits is more important feature than individual pixels~\cite{lee2009convolutional}. Therefore, instead of using the input features, we extract high level features more relevant to the classification task to detect vulnerable records. 

Specifically, when attacking neural networks (e.g., see Figure~\ref{fig:dnn} for a two-layer fully connected neural network), we construct new feature vectors by concatenating the outputs of the last layer before the Softmax function from the reference models (Figure~\ref{fig:features}), as the deeper layers in the network are more correlated with the classification output~\cite{hinton2006fast}. We then measure the unique influences of each record using its new feature vector. Let $\mathbf{f}$ be the the new feature vector of the record $r$. We call two records $r_1$ and $r_2$ {\em neighbors} if the cosine distance between their feature vectors $\mathbf{f}_1$ and $\mathbf{f}_2$ is smaller than a \emph{neighbor-threshold} $\delta$.

Note that the neighboring records are difficult to be distinguished by MIA because they have similar influence on the model. When a neighbor of $r$ occurs in the training dataset, the model may behave as if $r$ is used to train the model, leading to the incorrect membership inference result. Our goal is to select the vulnerable records in the entire record space with fewer or no neighbors likely to be present in the training set (assuming the training records are independently drawn from the record space) as putative targets of MIA.

Given a training dataset with $N$ records and a reference dataset with $N'$ records, both sampled from the same record space, and a target record $r$, we count the number of neighbors of $r$ in the reference dataset, denoted as $N'_n$. Then, the expected number of neighbors of $r$ in the training dataset, $N_n$, can be estimated as 
%
$\mathbb{E}\left[N_n\right] = N'_n \times \frac{N'}{N}.$

A record $r$ is considered to be potentially vulnerable (and as the attack object), only if $\mathbb{E}\left[N_n\right] < \beta$, where $\beta$ is the \emph{probability-threshold} for target record selection. We stress that the approach for vulnerable records selection presented here relies only on the record space (represented by the reference records accessible by an adversary) and the reference models (built using reference records), and is independent of the target model; as a result, the computation can be done off-line even when used to attack a machine learning as a service (MLaaS). 

\subsection{Direct Inference by Querying the Target Record}~\label{subsec:query_target}
In well-generalized models, a single record's influence on the model's prediction is usually small and hard to detect. Moreover, the extent of this influence varies between records, so the approach in the prior MIA~\cite{shokri2016membership} no longer works. Instead, we attack each target record separately by computing the deviation between its output given by the target model and those given by the reference models. 
We expect that each training record has a unique influence on the model, which can be measured by comparing the target model's output with the output of reference models (trained without the target record) on the record. 
We quantify the difference between the outputs using the log loss function. Given a classifier $M$ and a record $r$ with class label $y_r$, let $p_{y_r}$ be $M$'s output probability of class label $y_r$. The log loss function~\cite{murata1994network} $\mathcal{L}\left(M,r\right)$ is defined as:
$\mathcal{L}\left(M,r\right) = - \log p_{y_r}.$
The log loss function is commonly used as a criterion function~\cite{murata1994network} when training neural network models. $\mathcal{L}\left(M,r\right)$ is small when $M$ gives high probabilities on correct labels.

Given a target model $M$, a target record $r$, and $k$ reference models, we first obtain the log loss of all the reference models on $r$ as $L_1, L_2, \dots, L_k$. We view these losses as samples independently drawn from a distribution $\mathcal{D}(L)$, and estimate the empirical cumulative distribution function (CDF) of $\mathcal{D}_L$ as $F(L)$, which takes a real-valued loss $L$ as input.
We use the shape-preserving piecewise cubic interpolation~\cite{sprague1990shape} to smooth the estimated CDF. Based on the log loss of the target model $M$ on the target record $r$, $\mathcal{L}\left(M,r\right)$, we estimate the confidence of $r$ to be present in the training set by performing a left-tailed hypothesis test: under the null hypothesis $H_0$, $r$ is not present in the training set (i.e.,  $\mathcal{L}\left(M,r\right)$ is randomly drawn from $\mathcal{D}(L)$), while under the alternative hypothesis $r$ is used to train $M$ (i.e., $\mathcal{L}\left(M,r\right)$ is smaller than samples in $\mathcal{D}(L)$ because of the influence of $r$ in the training). Therefore, we calculate the $p$-value as:
%
$p=F\left(\mathcal{L}\left(M,r\right)\right),$
%
which gives the confidence that $r$ is used for training $M$ only if $p$  is smaller than a threshold (e.g. $0.01$) so that the null hypothesis is rejected.

\subsection{Indirect Inference without Querying the Target Record}
\begin{figure}[t]
\vspace{-30pt}
\centering
\includegraphics[width=0.5\textwidth]{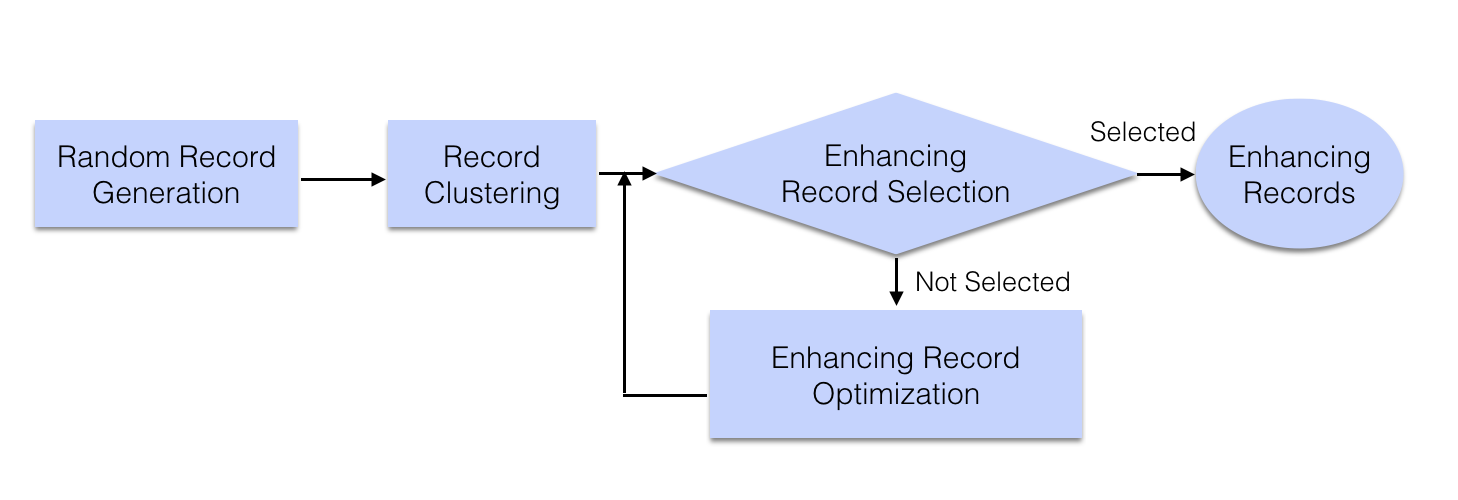}
\vspace{-20pt}
\caption{Steps for generating enhancing records.
\label{fig:multiQueryAttack}}
\vspace{-3pt}
\end{figure}
Besides reducing a model's loss on its own, a training record also influences the model's outputs on other records. This influence is desirable to improve model generalization: in order to give correct predictions on unseen records, a model needs to use the correlation it learns from a training record to make predictions on queries with similar features. On the other hand, however, these influences can be exploited by an adversary to obtain more information about the target record through multiple queries to enhance MIA. 
Interestingly, we show that MIA can be achieved by queries of records seemingly uncorrelated with the target record, making the attack hard to detect.

The key challenge for inference without querying the target record is to efficiently identify the {\em enhancing records} whose outputs from the target model are expected to be influenced by the target record. To address this problem, we develop a method consisting of the following steps: random record generation, record clustering, enhancing record selection, and enhancing records optimization (as shown in Figure~\ref{fig:multiQueryAttack}). 

\vspace{2pt}\noindent\textbf{Random Record Generation}. To start with, we randomly generate records from which the enhancing records are selected. Specifically, we adopt one of the following two methods for random record generation: (1) when the feature space is relatively small, we uniformly sample records from the whole feature space; (2) when the feature space is large, since the chance of getting enhancing records by uniform sampling is slim, we generate random records by adding Gaussian noise to pre-selected vulnerable target records. 

\begin{figure}[t]
\centering
\vspace{-20pt}
\includegraphics[width=0.5\textwidth]{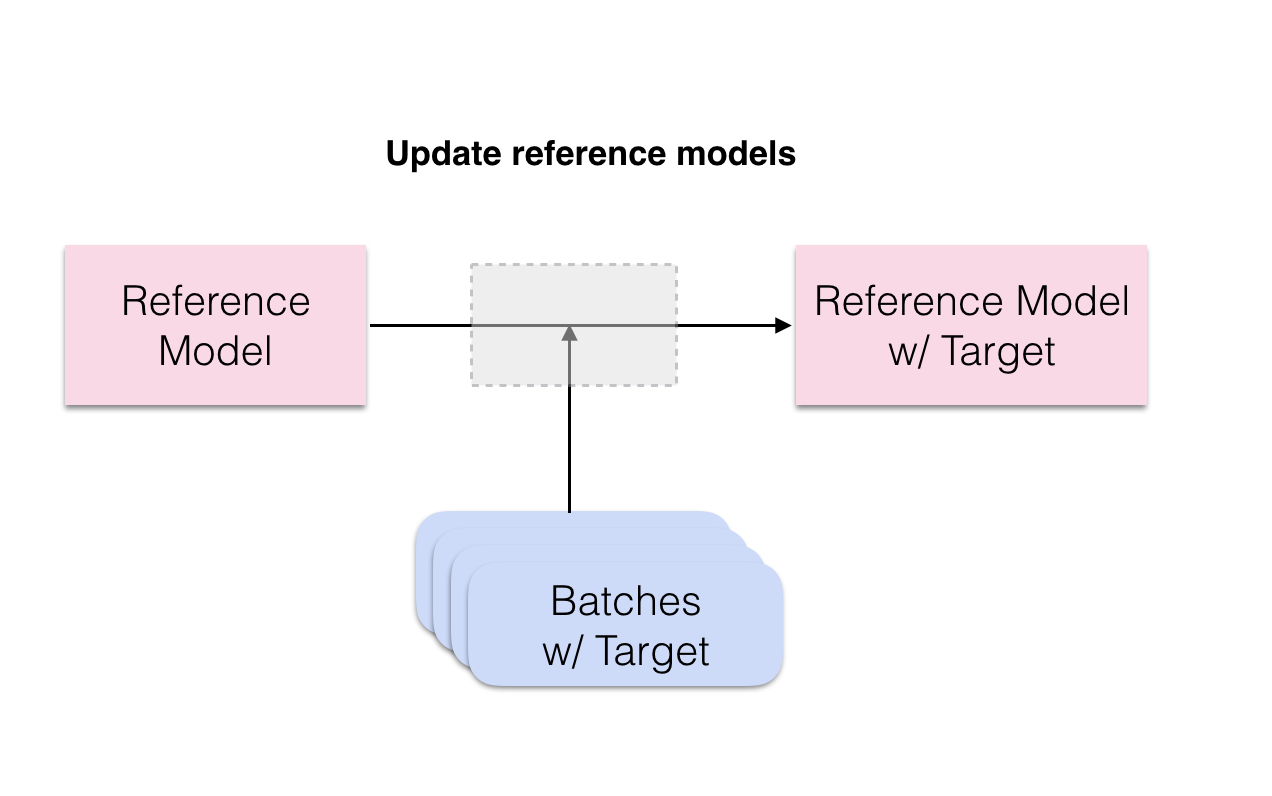}
\vspace{-30pt}
\caption{Building positive reference models by updating the model with the training set including the reference records plus the target record.
\label{fig:model_update}}
\end{figure}

\vspace{2pt}\noindent\textbf{Enhancing Record Selection}.
To identify records whose target model's output may be influenced by the target record $r$, we approximate the target model's behavior using a group of {\em positive reference models} that are trained using reference records plus the target record $r$. 
To save the effort of retraining the positive reference models, we add the target record into batches sampled from the original reference dataset and update the reference models by training on the batches plus the target record. Figure~\ref{fig:model_update} shows the process of updating reference models.

We select the {\em enhancing records} by comparing the predictions between the positive reference models (i.e., ``in models'') and the original reference models  (that are trained without the target records, i.e., ``out models''). We denote the $i$th original and the $i$th positive reference model as $M_{\textrm{ref}_i}$ and $M_{\textrm{ref}_i}^r$, respectively. 
Given a record $r$ with class label $y_r$ and another arbitrary record $q$, let $M\left(q,y_r\right)$ be the model $M$'s output probability of $y_r$ on the query $q$. We calculate $r$'s influence on $q$ as follows:
\begin{equation} \label{eq:influence}
	I\left(r,q\right) = \frac{1}{k} \sum_{i=1}^{k} \thresh\left(M_{\textrm{ref}_i}^r\left(q,y_r\right) -M_{\textrm{ref}_i}\left(q,y_r\right)\right) \ ,
\end{equation}
\noindent where $k$ is the total number of original (or positive) reference models, and $\thresh$ is a threshold function defined as follows:
\begin{equation*}
	\thresh\left(x\right) = \begin{cases}
		1 \qquad \textrm{if} \quad  x>0, \\
        0 \qquad \textrm{otherwise}.
	\end{cases}
\end{equation*}

\begin{algorithm}[h]
\caption{Enhancing Records Selection Algorithm\label{alg:query_sel}}
\begin{algorithmic}[1]
\Procedure{${\rm select}_{\theta}$}{$q$}\Comment{Input a random query}
\State $I\left(r,q\right) \gets \sum_{i=1}^{k} \thresh\left(M_{\textrm{ref}_i}^r\left(q,y_r\right) -M_{\textrm{ref}_i}\left(q,y_r\right)\right)\big/k$ 
\If{$I > \theta$} 
\State Accept $q$ \Comment{Use $q$ in MIA}
\Else
\State Reject $q$
\EndIf
\EndProcedure
\end{algorithmic}
\vspace{2pt}
\end{algorithm}

We identify a randomly generated record $q$ is an enhancing record for the record $r$ if $I(r,q)$ approaches 1, which indicates that adding $r$ to the training dataset \emph{almost always} increase the models' output probability on the class label $y_r$ for the query $q$. In practice, we use $q$ in the MIA on the target record $r$ only if $I(r,q)$ is greater than a threshold $\theta$ (e.g. 0.95). Algorithm~\ref{alg:query_sel} summarizes the entire algorithm for query selection.

\vspace{2pt}\noindent\textbf{Enhancing Record Optimization}. 
When the target model has a large record space (e.g., with high-dimension feature vectors), the chance of finding an enhancing record  among randomly generated records is slim. To address this issue, we propose an algorithm to search for enhancing records for a target record $r$ by optimizing the following objective function:
\begin{equation} \label{eq:obj_inf}
	\max_{q}I\left(r,q\right),
\end{equation}
\noindent where $I\left(r,q\right)$ is the influence function defined in Equation~\ref{eq:influence}. Optimizing $I\left(r,q\right)$ is time-consuming because 
$I\left(r,q\right)$ consists of a non-differentiable threshold function $\thresh$. Therefore, instead of solving the optimization function in equation~\ref{eq:obj_inf}, 
For simplification, we approximate the maximization of $I\left(r,q\right)$ with the minimization of the sum of multiple hinge loss functions defined as follows~\cite{gentile1999linear}:
\begin{equation}\label{eq:obj_loss}
	\min_{q}\sum_{i=1}^{k}\max\left(0, \gamma - \left(M_{\textrm{ref}_i}^r\left(q,y_r\right) -M_{\textrm{ref}_i}\left(q,y_r\right)\right)\right),
\end{equation}
\noindent where $\gamma$ is a parameter indicating the margin width. 
If a randomly generated record are rejected by the query selection algorithm, we minimize the objective function in Equation~\ref{eq:obj_loss} using gradient descent~\cite{cauchy1847methode} to check if the resulting record is acceptable as an enhancing record. 

\vspace{2pt}\noindent\textbf{Record Clustering (Optional)}. 
Note that it is inefficient to repeat the query selection and optimization algorithms on all random records because the predictions of the models on most records are highly correlated: the models giving high output probabilities on some record are also likely to give high output probabilities on correlated records. To improve the efficiency of query selection, we propose an algorithm to identify the {\em least correlated} enhancing records from a large number of randomly generated records.

\begin{figure}[t]
\centering
\vspace{-20pt}
\includegraphics[width=0.5\textwidth]{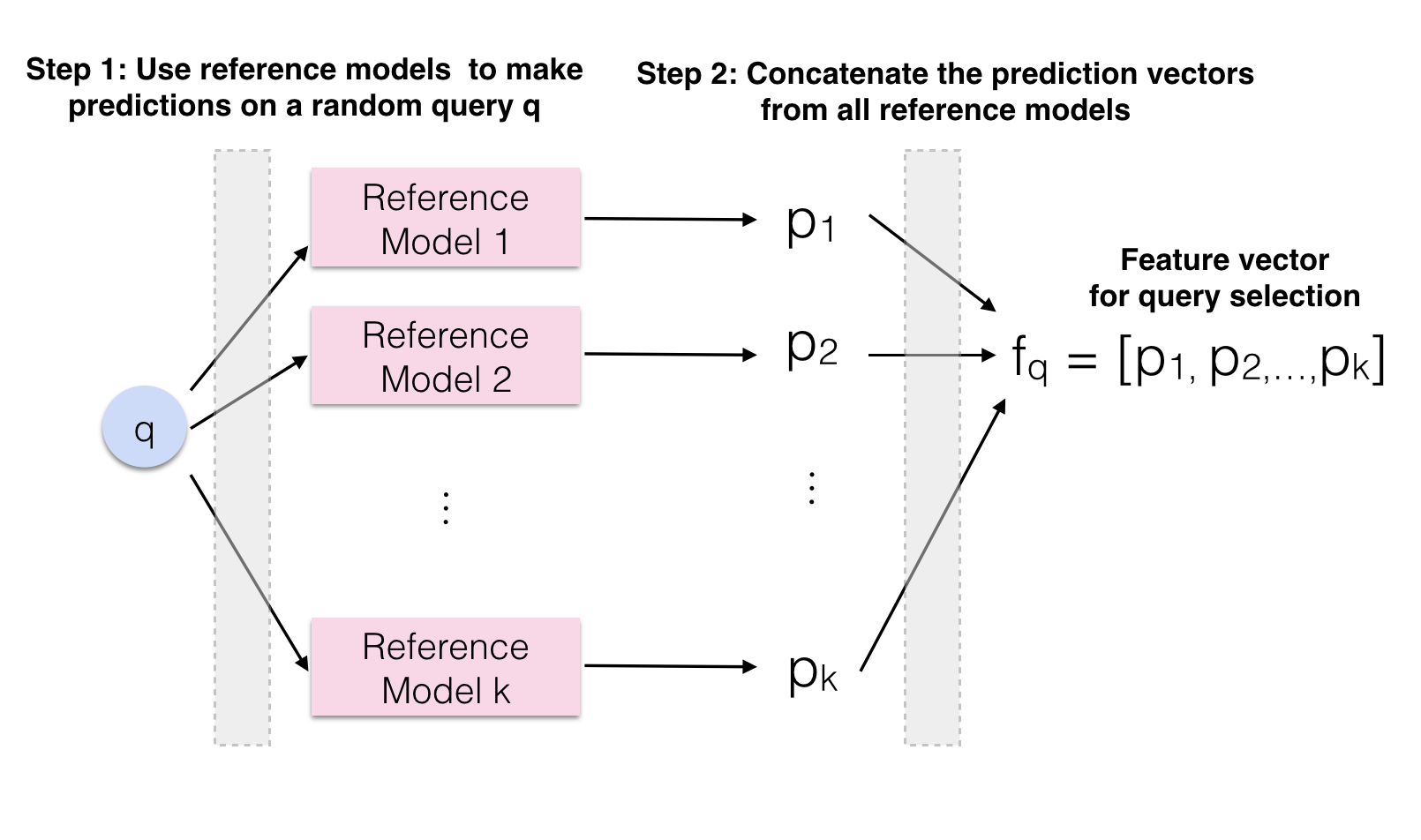}
\vspace{-20pt}
\caption{Generate query features for query selection.
\label{fig:query}}
\end{figure}

First, we estimate the correlation between records based on the model's predictions on them. We construct a feature vector $\mathbf{f}_q$ for a record $q$ by concatenating the reference models' outputs on it (Figure~\ref{fig:query}). If two queries $q_1$ and $q_2$ have highly correlated feature vectors, the models' outputs on $q_2$ do not add much information to the models' outputs on $q_1$. 

Next, we formulate the problem of selecting a subset of least correlated records as a graph theoretical problem. We build a graph where records are the nodes and pairwise correlation between records is the weight on edges connecting the corresponding nodes. This allows us to recast our problem as the $k$-lightest subgraph problem~\cite{watrigant2012k}, which is NP-hard.
We obtain an approximate solution using hierarchical clustering~\cite{jain1988algorithms}. For this, we cluster the records into $k$ disjoint clusters based on their pairwise cosine distance. Finally, in each cluster, we select the record with least average cosine distance to all other records in the same cluster. 

As shown in Figure~\ref{fig:multiQueryAttack}, we use the enhancing record clustering algorithm before the enhancing record selection and enhancing record optimization steps to improve the efficiency of the attack. 

\vspace{2pt}\noindent\textbf{Indirect Inference with Multiple Queries}. After identifying multiple enhancing records, we repeat the attack in section~\ref{subsec:query_target} by querying each of these records. Because the outputs on these queries may be correlated, we combine the resulting $p$-values using Kost's method~\cite{kost2002combining}, with the covariance matrix estimated from the query features generated in the query selection step (Figure~\ref{fig:query}).

\section{Evaluation}
\label{sec:eval}

\subsection{Experimental Setup}
We evaluated two aspects of the performance of our attack: (1) \textit{How many target records are considered to be vulnerable according to the GMIA selection criterion?} and (2) \textit{How likely are vulnerable records to be inferred by GMIA when they are in the training dataset?}

To answer the first question, we ran the GMIA vulnerable record selection algorithm over all the target records. We compared the number of selected vulnerable records across different datasets, varying neighbor threshold $\delta$, and probability threshold $\beta$. We evaluated the performance of GMIA over the selected vulnerable target records instead of the whole dataset since, in real attacks, adversary is likely to choose a few vulnerable targets instead of attacking all individuals.

To answer the second question, we evaluated the performance of the attack over multiple models. We constructed $100$ target models, half of which are trained with the target record. To guarantee that each target record occurred in exactly $50$ out of $100$ target models, we generated training datasets by randomly splitting the target records into two datasets of the same size, each serving as a training set for a target model. We repeated this process for $50$ times and generated the training datasets for $100$ target models.

For each vulnerable target record, we performed GMIA on all the target models, and calculated for what percentage of models it can be correctly identified. When there are multiple vulnerable target records, we repeated the attack on every vulnerable target record over all the target models. An inference takes place only if the adversary has high confidence in the success of the attack (e.g. $p<0.01$). The {\em precision} of the attack is defined as the percentage of successful inferences (i.e., the target record is indeed in the training dataset) among all inferences. The {\em recall} of the attack is defined as the percentage of successful inferences among all the cases that the target record is in the training set (i.e. $50n$). It indicates the likelihood that the membership of a vulnerable target record can be inferred. We define {\em true positive (TP)} to be the case that the target record is indeed in the training dataset when the adversary inferred it as in and {\em false positive (FP)} to be the case that the target record is {\em not} in the training dataset when the adversary inferred it as in. 

\subsection{Dataset}

\noindent\textbf{UCI Adult}. The UCI Adult dataset~\cite{Lichman:2013} is a census dataset containing 48,842 records and 14 attributes. The attributes are demographic features and the classification task is to predict whether an individual's salary is above \$50K a year. We normalized the numerical attributes in the dataset and used one hop encoding~\cite{wu2006local} to construct the binary representation of categorical features. We randomly selected 20,000 records for training target models, and each training dataset contains 10,000 records. The remaining 28,842 records served as the adversary's background knowledge.

\vspace{3pt}\noindent\textbf{UCI Cancer}. The UCI cancer dataset~\cite{Lichman:2013} contains 699 records and 10 numerical features ranging between 1 to 10. The features are characteristics of the cell in an image of a fine needle aspirate (FNA) of a breast mass. The classification task is to determine whether the cell is malignant or benign. We randomly selected 200 records for training, and each training dataset contains 100 records. The remaining 499 records served as the adversary's background knowledge.

\vspace{3pt}\noindent\textbf{MNIST Dataset}.
The MNIST dataset~\cite{lecun2010mnist} is an image dataset of handwritten digits. 
The classification task is to predict which digit is represented in an image. We randomly selected 20,000 images for training and 40,000 images as the adversary's background knowledge. Each training set for target models and reference models contains 10,000 images. 

\subsection{Models}

\noindent\textbf{Neural Network}.
For the Adult dataset, we constructed a fully connected neural network with 2 hidden layers with 10 units and 5 units respectively. We use \texttt{Tanh} as the activation function and \texttt{SoftMax} as the output layer. The model is trained with batchsize of 100 and 20,000 epochs. For the MNIST dataset, we constructed 2 convolutional layers with ReLu as the activation function, followed with max pooling layers. We then added a fully connected layer of 1,024 neurons, and we also used dropout techniques to reduce overfitting. Finally, we added an output layer and a Softmax layer. The model is trained with batchsize of 50 and 10,000 epochs. For the Cancer dataset, we used a vanilla neural network with no hidden layer. The model is trained with batchsize of 10 and 3,000 epochs.

\vspace{3pt}\noindent\textbf{Google ML Engine}. Since the Google Predictions API~\cite{googlePredApi} used in the prior attack is deprecated, we used Google ML Engine~\cite{googleMl} to train target models on ML cloud. When training the model, we used the  samples provided by Google, which has pre-built model structures for training models on Adult dataset and MNIST dataset. Specifically, for Adult dataset, the sample code uses Google estimator~\cite{abadi2016tensorflow} which hides low-level model structure from the user; for MNIST dataset, the sample code builds a neural network with 2 fully-connected hidden layers.



\begin{table}[!t]
{\scriptsize
\centering
\vspace{-10pt}
\caption{\fontsize{9}{11}\selectfont GMIA by Direct Inference}
\label{tab:direct}
\begin{tabular}{C{1.5cm}C{1.25cm}C{1.25cm}C{1.25cm}C{0.5cm}C{0.5cm}}
\hline
& $p$-value & precision & recall & TP & FP \\ 
\hline
\multirow{3}{*}{\specialcell{Adult \\ ($13$ records)}}
 & $0.001$ & - & $0$ & $0$ & $0$ \\
 & $0.002$ & 1 & $0.31\%$ & $2$ & $0$ \\
 & $0.01$ & $73.91\%$ & $5.23\%$ & $34$ & $12$ \\
\hline
\multirow{3}{*}{\specialcell{Cancer \\ ($5$ records)}} 
 & $0.001$ & - & $0$ & $0$ & $0$ \\
 & $0.008$ & 1 & $2.4\%$ & $6$ & $0$ \\
 & $0.01$ & $88.89\%$ & $3.2\%$ & $8$ & $1$ \\
 \hline
\multirow{3}{*}{\specialcell{MNIST \\ ($16$ records)}}
 & $0.0001$ & $1$ & $24.37\%$ & $195$ & $0$ \\
 & $0.001$ & $96.55\%$ & $45.50\%$ & $364$ & $13$\\
 & $0.01$ & $93.36\%$ & $73.88\%$ & $591$ & $42$ \\
\hline
\multirow{3}{*}{\specialcell{Adult(Google) \\ ($7$ records)}}
 & $0.001$ & - & $0$ & $0$ & $0$ \\
 & $0.009$ & $1$ & $2.33\%$ & $7$ & $0$ \\
 & $0.01$ & $80\%$ & $2.67\%$ & $8$ & $2$ \\
\hline
\multirow{3}{*}{\specialcell{MNIST(Google) \\ ($1$ record)}} & $0.001$ & - & $0$ & $0$ & $0$ \\
 & $0.01$ & $1$ & $4\%$ & $2$ & $0$ \\
 & $0.014$ & $1$ & $8\%$ & $4$ & $0$ \\
\hline
\end{tabular}
}
\end{table}

\subsection{Direct Inference}

In our first attack, we inferred the membership of vulnerable target records from the target models' predictions on these records. Based on the vulnerable target record selection algorithm in Section~\ref{sec:attack}, using a probability threshold $\beta = 0.1$ (i.e. the likelihood that a target record's neighbor occurs in the training dataset of the target model is smaller than $0.1$) , we selected $13$ (out of $20,000$) target records in the Adult dataset, $5$ (out of $200$) target records in the Cancer dataset, and $16$ (out of $20,000$) target records in the MNIST dataset. The neighborhood threshold $\delta$ used for these three datasets are $0.4$, $0.1$, and $0.2$ respectively. We discuss the influence of these parameters in Section~\ref{subsec:target}. For models trained on Google ML engine, we selected $7$ (out of $20,000$) target records for Adult(Google) and $1$ target record for MNIST(Google). We performed GMIA on each of the selected target record and on all $100$ target models. 

\begin{figure}
	\vspace{-10pt}
    \centering
	\includegraphics[width=0.45\textwidth]{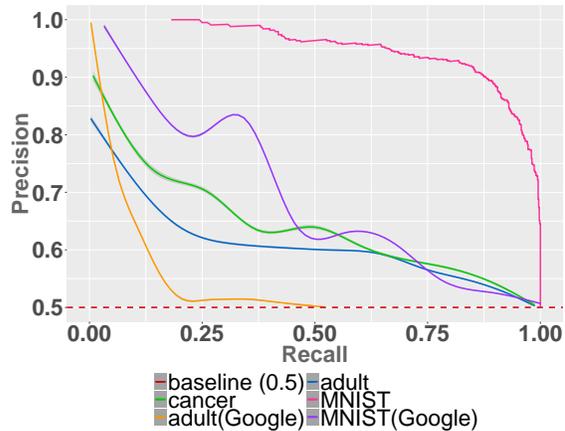}   
    \caption{GMIA by Direct Inference}\label{fig:direct-rp}
\end{figure}

Figure~\ref{fig:direct-rp} shows the precision-recall curve of GMIA by querying the target record. Table~\ref{tab:direct} reflects the attack performance under different cut-off $p$-values. The recall reflects the likelihood that the membership of selected target records will be identified. When using $0.01$ cut-off threshold for $p$-values, an adversary can attack with $73.91\%$ precision on the Adult dataset, $88.89\%$ precision on the Cancer dataset, and $93.36\%$ precision on MNIST. All the target models we successfully attacked are well-generalized with difference between training and testing accuracy below $0.01$(Table~\ref{tab:acc} in the Appendix). In comparison, the prior MIA~\cite{shokri2016membership} has low precision ($<70\%$) on the same models and the same target records as shown in Table~\ref{tab:prior_attack} in the Appendix. 

Our attack had better performance on the local MNIST model compared to the Google ML ones because the CNN we constructed locally was more complex. Note that our local CNN improved upon the model on Google ML engine in testing accuracy by $8\%$, indicating an increase in model utility. However, the privacy risk also increased significantly. When $p<0.01$, the attack recall increased by more than $70\%$. This result indicates the high privacy risk of applying complex models even when these models are not overfitted. 

Our vulnerable record selection mechanism was less effective on the Adult Google ML model since we did not have access to the exact model structure due to the use of Google estimator. Instead, we used raw features to select target records. This limitation reduced the number of vulnerable target records we identified from $13$ to $7$. 

\begin{figure*}
    \centering
    \vspace{-30pt}
     \begin{subfigure}[b]{0.3\textwidth}
     \includegraphics[width=\textwidth]{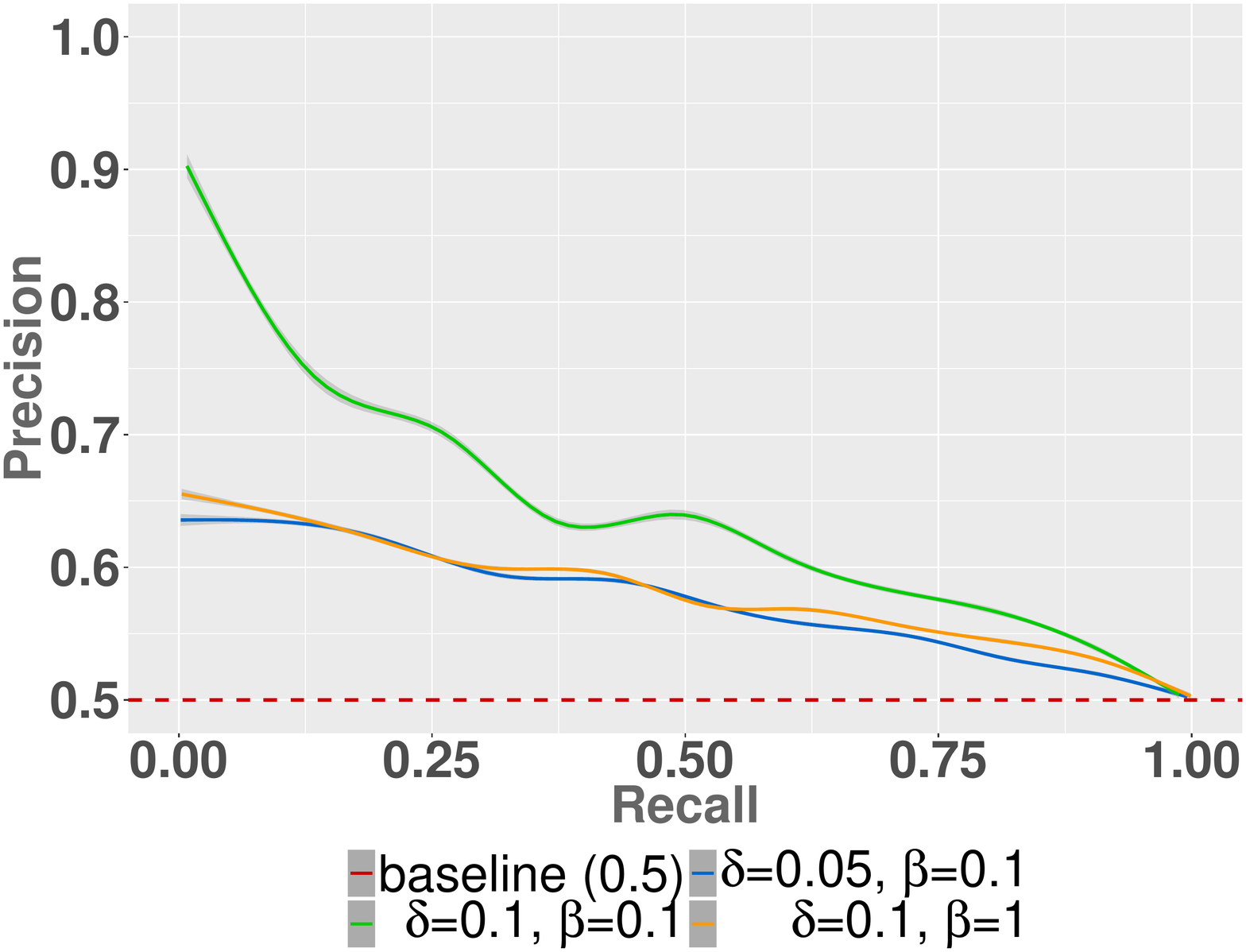}
        \caption{\footnotesize Cancer Dataset}
        \label{fig:cancer-target-rp}
    \end{subfigure}    
    \begin{subfigure}[b]{0.3\textwidth}
\includegraphics[width=\textwidth]{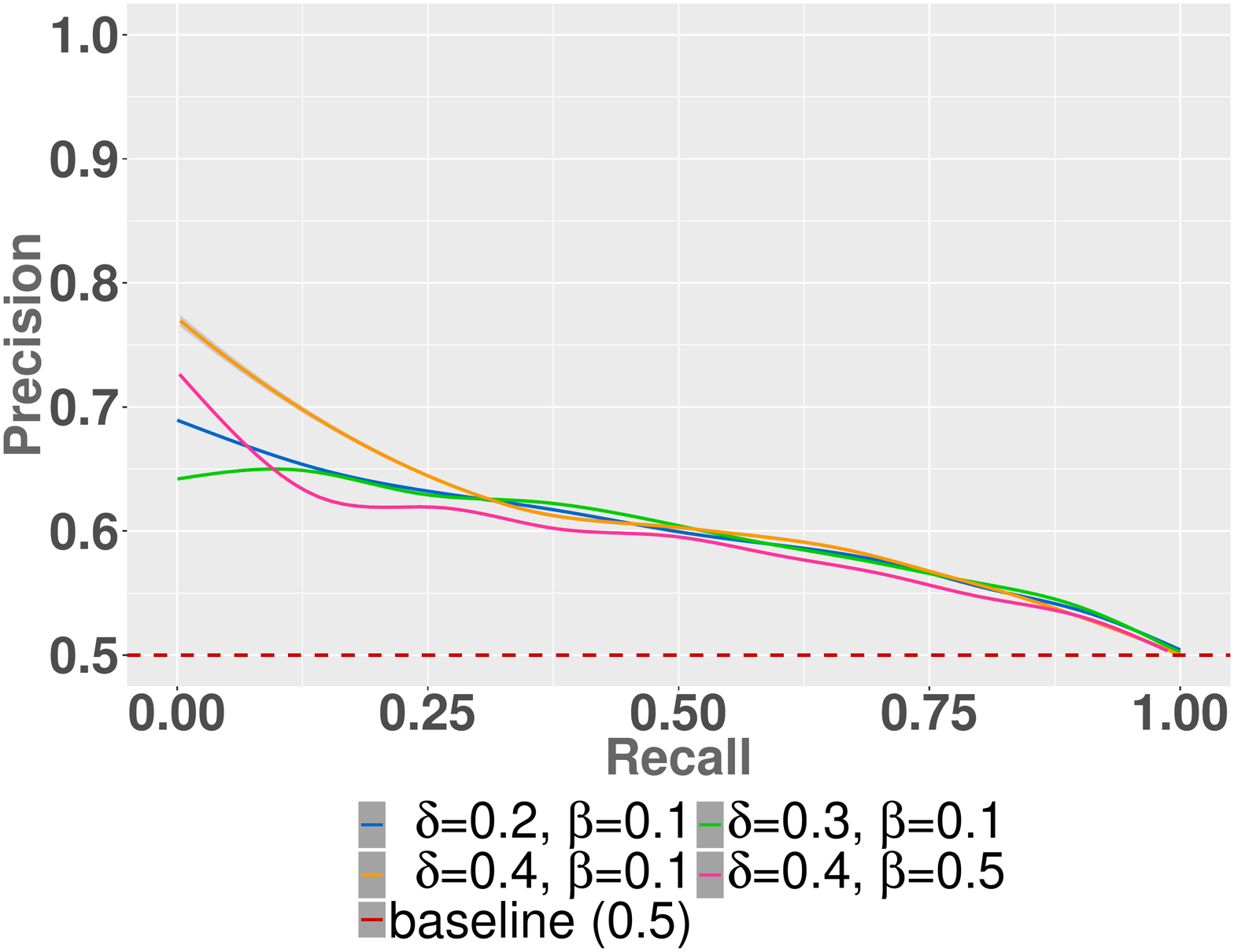}
        \caption{\footnotesize Adult Dataset}
        \label{fig:adult-target-rp}
    \end{subfigure}
    \begin{subfigure}[b]{0.3\textwidth}
        \includegraphics[width=\textwidth]{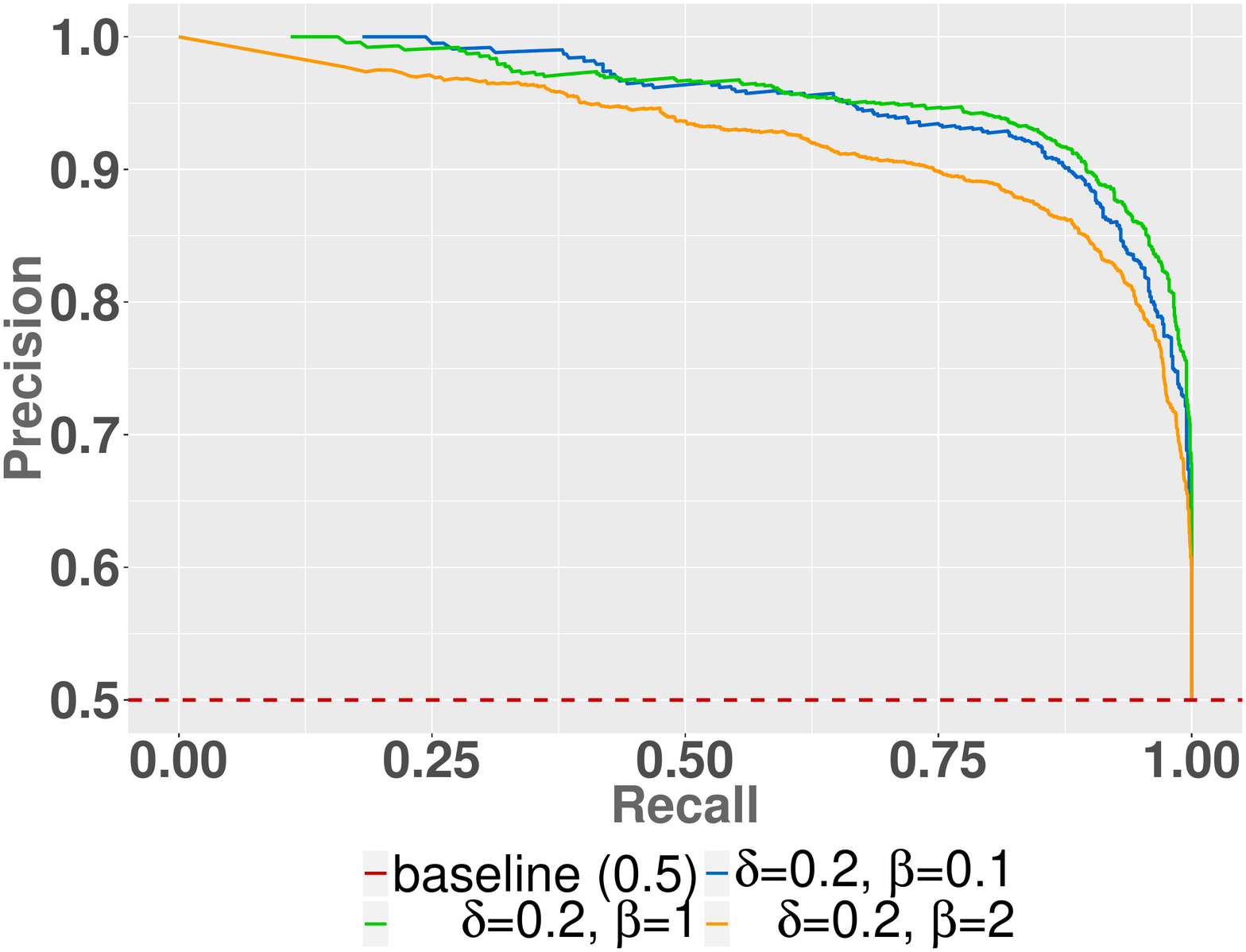}
        \caption{\footnotesize MNIST Dataset}
        \label{fig:mnist-target-rp}
    \end{subfigure}
    \caption{Effect of different target selection threshold on attack performances}\label{fig:target}
    \vspace{-10pt}
\end{figure*}

\subsection{Influence of Vulnerable Target Record Selection~\label{subsec:target}}
Before launching the attack, we selected vulnerable target records by finding out records with unique high level feature vectors. This selection process helps reducing the incorrect inference caused by similar records in the training dataset. The selection criterion depends on two parameters: the neighbor threshold $\delta$, which determines the criterion of neighbors, and the probability threshold $\beta$, which indicates how likely a neighbor is to occur in the training dataset. We studied selected vulnerable target records under different thresholds. 
Table~\ref{tab:target} and Figure~\ref{fig:target} shows performance of GIA w.r.t. varying target record selection threshold. Smaller neighbor thresholds or higher probability thresholds increased the number of selected vulnerable target records. However, as we tried to attack more records at the same time, there was a higher chance that we would make false positive inferences due to the influence of a record similar to one of the target records, which decreased the attack precision. Moreover, the recall of the attack also decreased since we included records with weaker influence on the model as vulnerable target records.

\begin{table}[t] 
{\scriptsize
\centering
\caption{\fontsize{9}{11}\selectfont GMIA w.r.t. Target Record Selection ($p=0.01$)}
\label{tab:target}
\begin{tabular}{C{1.5cm}C{0.5cm}C{0.5cm}C{1.5cm}C{1.5cm}C{1.5cm}}
\hline
& $\delta$ & $\beta$ & \# of Targets & precision & recall \\ \hline
\multirow{4}{*}{Adult} & $0.4$ & $0.1$ & $13$ & $73.91\%$ & $5.23\%$\\
 & $0.4$ & $0.5$ & $26$ & $68\%$ & $3.92\%$ \\
 & $0.3$ & $0.1$ & $53$ & $65.81\%$ & $2.91\%$ \\
 & $0.2$ & $0.1$ & $127$ & $66.21\%$ & $2.19\%$ \\
 \hline
\multirow{3}{*}{Cancer} & $0.1$ & $0.1$ & $5$ & $88.89\%$ & $3.2\%$\\
 & $0.1$ & $1$ & $21$ & $68.75\%$ & $3.14\%$ \\
 & $0.05$ & $0.1$ & $33$ & $66.67\%$ & $2.6\%$ \\
\hline
\multirow{3}{*}{MNIST} & $0.2$ & $0.1$ & $16$ & $93.36\%$ & $73.88\%$\\
 & $0.2$ & $1$ & $27$ & $95.05\%$ & $66.89\%$ \\
 & $0.2$ & $2$ & $52$ & $90.84\%$ & $68.31\%$ \\
 \hline
\end{tabular}
}
\end{table}

\subsection{Indirect Inference}

For some vulnerable target records, we achieved the same level of attack performance by querying enhancing records. For each dataset, we randomly sampled $5,000$ records, selected $50$ of them by record clustering, and tested them with the enhancing record selection algorithm~\cite{berkhin2006survey}. If less than $10$ enhancing records were selected, we ran the enhancing record optimization algorithm to improve the records. The initial records for the Cancer dataset and the Adult dataset were randomly sampled from the feature space while the records for the MNIST dataset were generated by adding noise to the target records due to the large feature space. 

We selected 1 target record in each dataset. For the Cancer dataset, we selected $47$ enhancing records whose euclidean distance to the target record range between $6$ and $19.3$ with a selection criterion $I\left(r,q\right)>0.95$. Since the Cancer dataset has relatively low dimensional features, enough enhancing records were accepted, and enhancing record optimization was not needed. For the Adult dataset, we relaxed the enhancing record selection criterion to $I\left(r,q\right)>0.9$ and found $15$ enhancing records after the optimization step. For the MNIST dataset, we further relaxed the enhancing record criterion to $I\left(r,q\right)>0.8$ due to the high dimensional feature space. We identified $41$ enhancing records generated by adding noise to the target record. 

\begin{figure*}
    \centering
    \vspace{-30pt}
     \begin{subfigure}[b]{0.3\textwidth}
    \includegraphics[width=\textwidth]{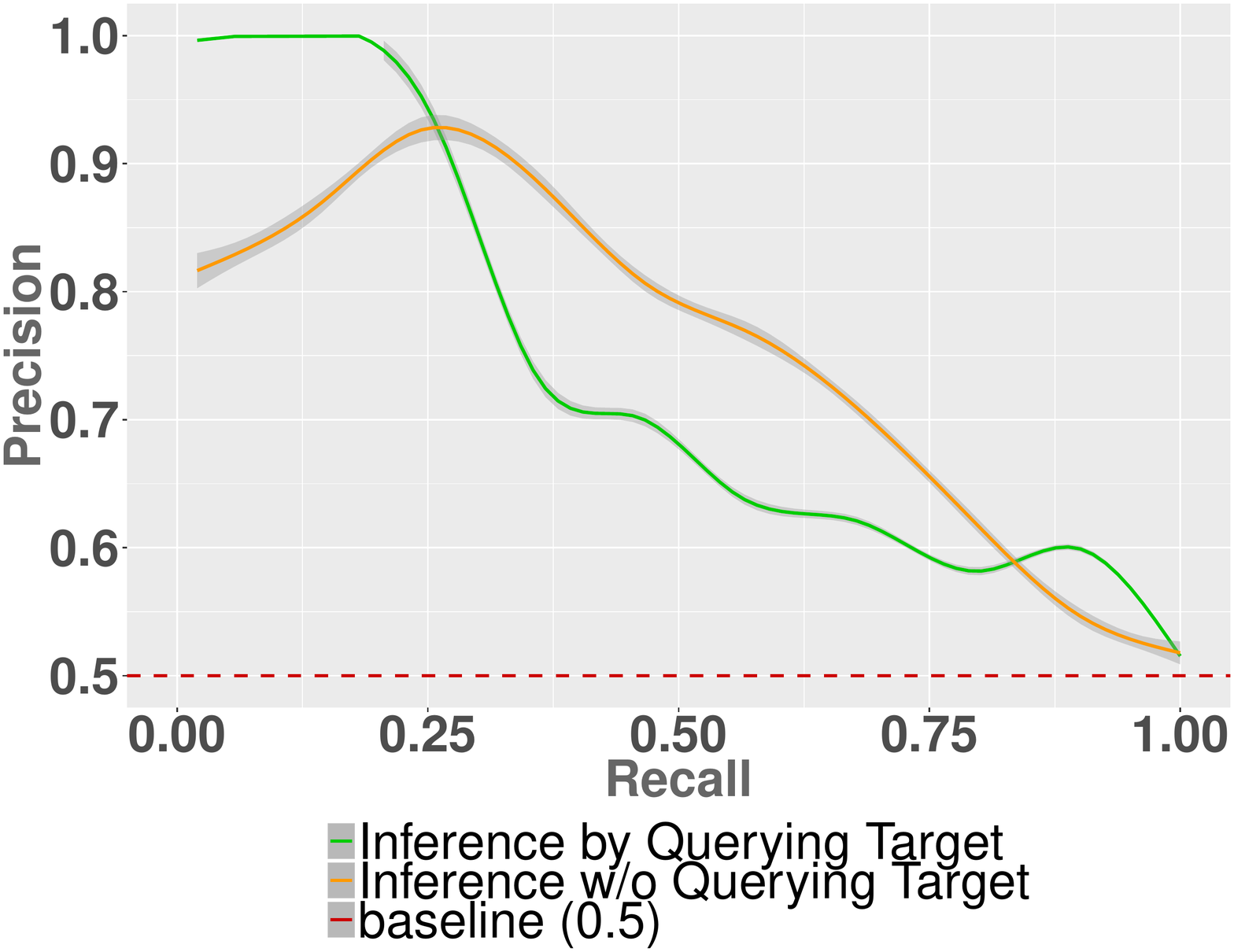}   
    \caption{Cancer Dataset}
    \label{fig:cancer-indirect}
    \end{subfigure}    
    \begin{subfigure}[b]{0.3\textwidth}
\includegraphics[width=\textwidth]{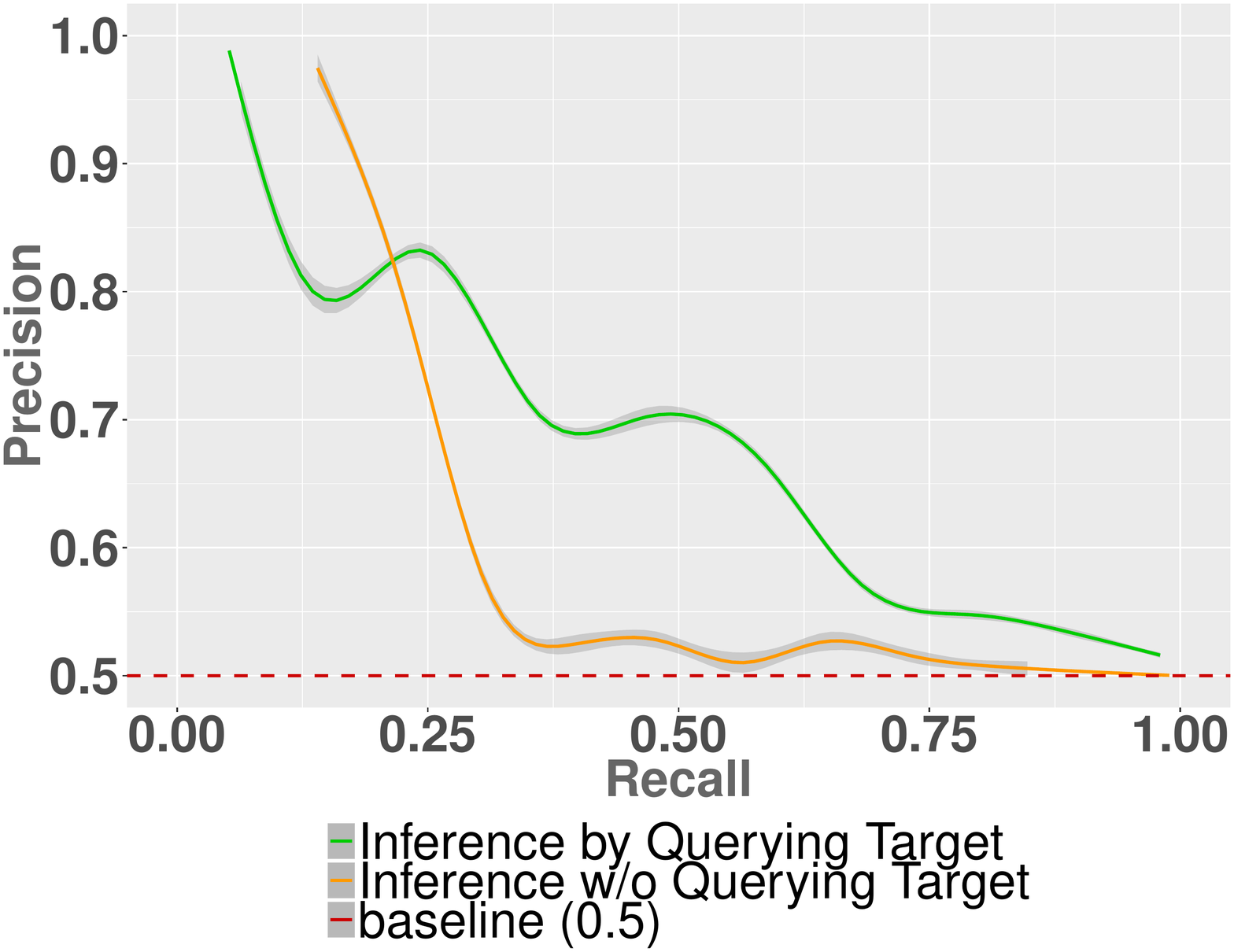}   
    \caption{Adult Database}
    \label{fig:adult-indirect}
    \end{subfigure}
    \begin{subfigure}[b]{0.3\textwidth}
        \includegraphics[width=\textwidth]{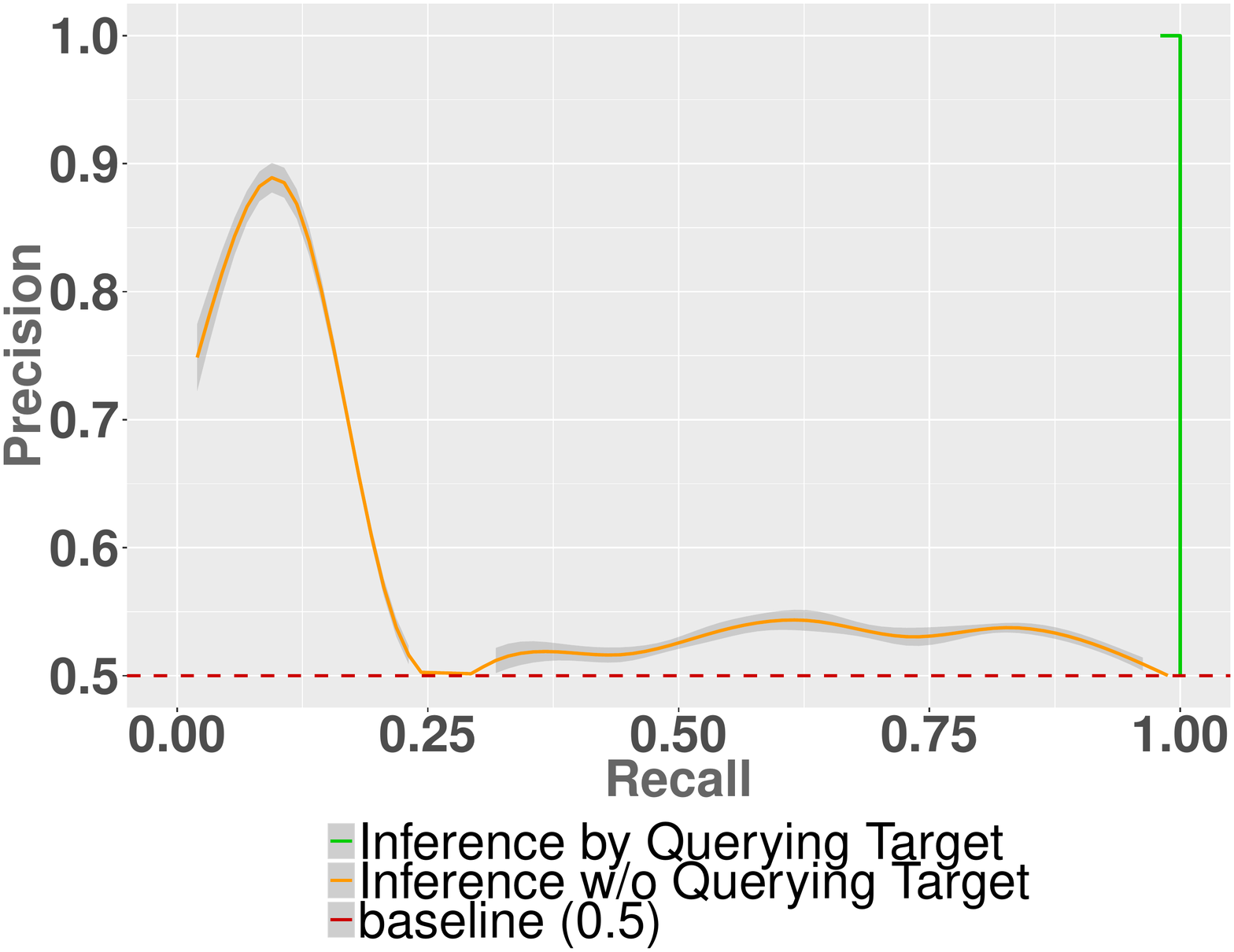}
        \caption{MNIST Dataset}
        \label{fig:mnist-indirect-rp}
    \end{subfigure}
    \caption{Attack without Querying the Target Record}\label{fig:indirect}
    \vspace{-10pt}
\end{figure*}

Table~\ref{tab:indirect} and Figure~\ref{fig:indirect} show the performance of indirect inferences. For both the Cancer dataset and the Adult dataset, attacking with the enhancing records has compatible performance as querying the target record. Moreover, for the Adult dataset, querying the target record did not successfully infer any cases with a $0.01$ cut-off $p$-value, but by combining the predictions on enhancing records, we achieved a precision of $1$ and a recall of $14\%$. 
For the MNIST dataset, we achieved a precision of $1$ and a recall of $2\%$ when $p\leq 0.01$. Although this performance is less impressive compared to a direct inference on the same record (whose precision and recall are both close to $1$) it's still an indication that membership inference attack can succeed without querying the target record. Moreover, we plotted both the target record and the enhancing records and found that the enhancing records in no means represent the target record, indicating that GMIA is hard to detect (Figure~\ref{fig:vulneralbe_record_mnist} in the Appendix).

\begin{table}[t]
{\scriptsize
\centering
\caption{\fontsize{9}{11}\selectfont Comparison between Direct and Indirect Inferences}
\label{tab:indirect}
\centering
\scriptsize{
\begin{tabular}{C{1.5cm}C{1.5cm}C{1cm}C{1cm}C{1cm}C{1cm}C{1cm}}
\hline
Dataset & $p$-value & prec. (direct) & recall (direct) & prec. (indirect) & recall (indirect) \\ 
\hline
 \multirow{2}{*} {\specialcell{Adult}}
 & $0.01$ & - & 0 & $1$ & $14\%$ \\
 & $0.1$ & $70.83\%$ & $34\%$ & $75\%$ & $24\%$ \\
 \hline
 \multirow{2}{*} {\specialcell{Cancer}}
 & $0.01$ & $1$ & $6\%$ & - & $0$ \\
 & $0.1$ & $66.67\%$  & $52\%$ & $88.89\%$ & $16\%$ \\
 \hline
 \multirow{2}{*} {\specialcell{MNIST}}
 & $0.01$ & $96.15\%$ & $1$ & $1$ & $2\%$  \\
 & $0.1$ & $89.29\%$ & $1$ & $52.38\%$ & $22\%$  \\
\hline
\end{tabular}}
}
\end{table}

\begin{figure*}
    \centering
     \begin{subfigure}[b]{0.3\textwidth}
    \includegraphics[width=\textwidth]{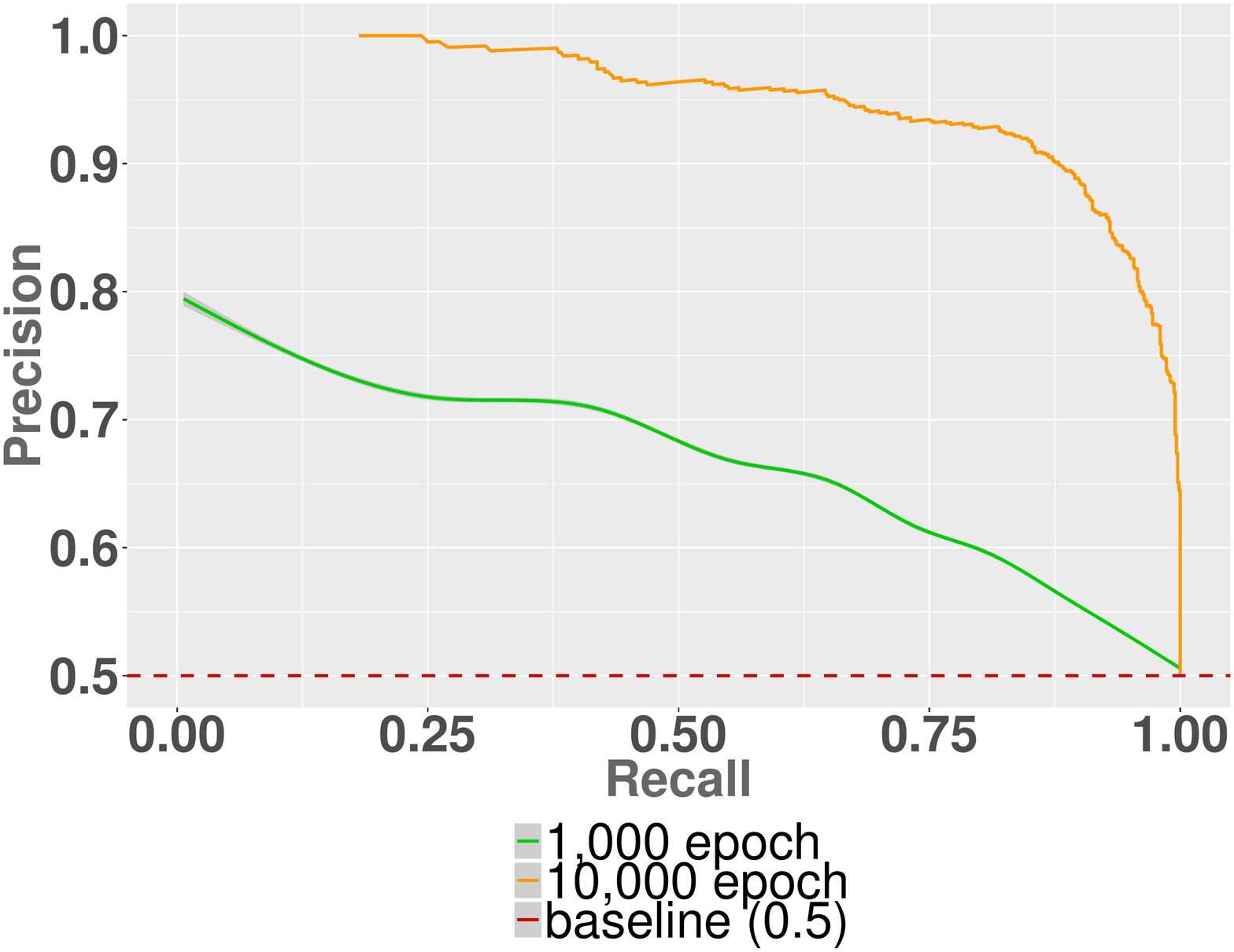}   
    \caption{\footnotesize GMIA Performance w.r.t Training epoch}
    \label{fig:mnist-epoch-rp}
    \end{subfigure}    
    \begin{subfigure}[b]{0.3\textwidth}
\includegraphics[width=\textwidth]{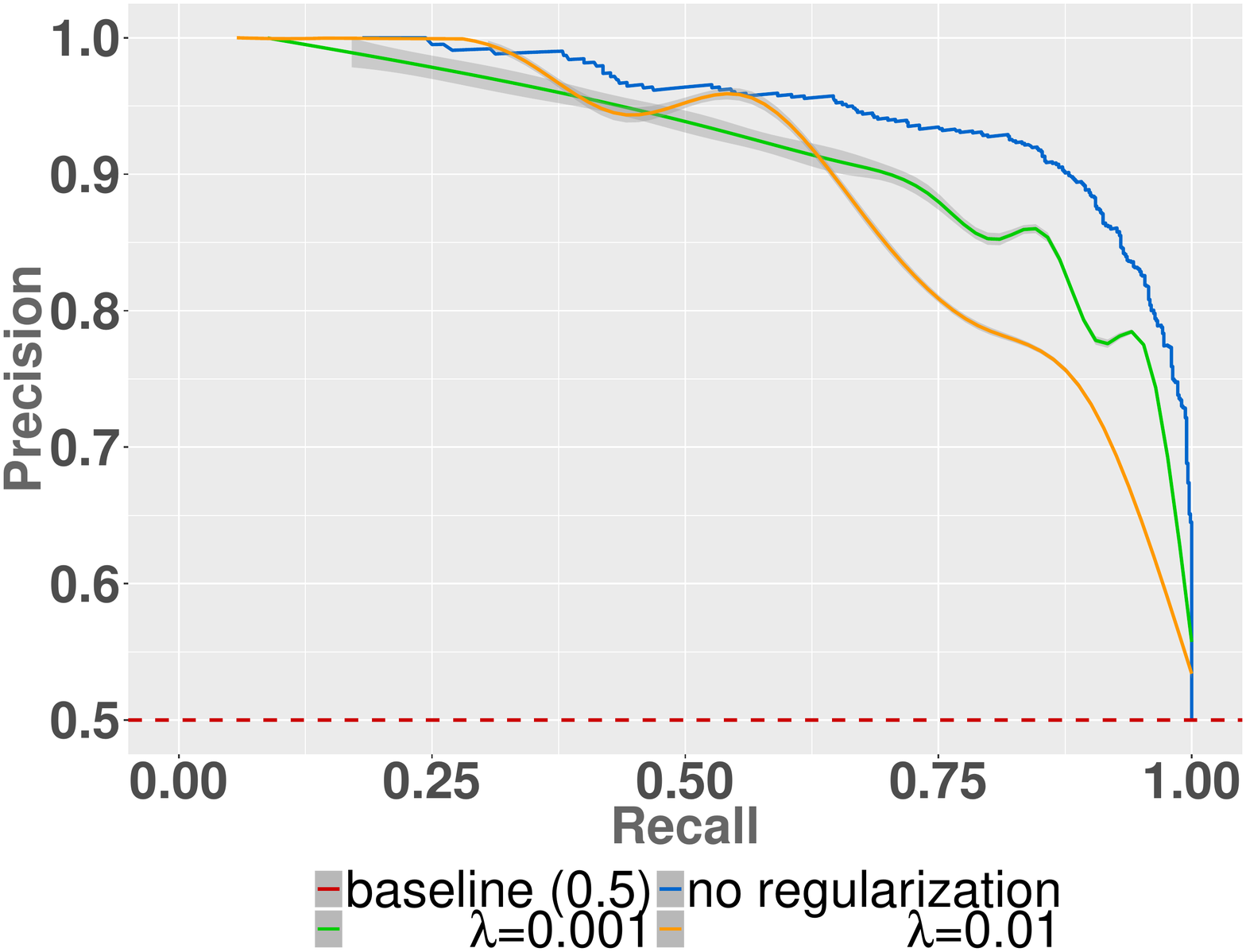}   
    \caption{\footnotesize GMIA Performance w.r.t. Regularization}
    \label{fig:mnist-l2}
    \end{subfigure}
    \caption{Influence of training epochs and regularization}
    \vspace{-20pt}
\end{figure*}

\subsection{Influence of Training Epochs}
In machine learning, one way of preventing overfitting is to stop training the model as soon as the testing accuracy stops increasing~\cite{caruana2001overfitting}. This method is called ``early stop''. To study the influence of maximum training epochs on GMIA, we trained neural networks on MNIST dataset with $1$k maximum training epochs. Unlike ``early stop'' method, which stops the training process after testing accuracy stops increasing, we stopped training the models {\em before} the testing accuracy stopped increasing and performed the attack on potentially underfitted models. Table~\ref{tab:epoch} shows the training and testing accuracy of the models

\begin{table}[h]
{\scriptsize
\vspace{5pt}
\centering
\caption{\fontsize{9}{11}\selectfont GMIA w.r.t. Training Epochs ($p=0.01$)}
\label{tab:epoch}
\begin{tabular}{C{1.1cm}C{1.1cm}C{0.6cm}C{0.4cm}C{0.4cm}C{1cm}C{1cm}C{1.1cm}}
\hline
Training Epoch & Training Acc. & Test Acc. & $\delta$ & $\beta$ & \# of Targets & prec. & recall \\
\hline
\multirow{2}{*}{$1,000$} & \multirow{2}{*}{$0.97$} & \multirow{2}{*}{$0.96$} & $0.2$ & $0.1$ & $28$ & $72.27\%$ & $6.14\%$ \\
& & & $0.3$ & $0.1$ & $4$ & $1$ & $2.5\%$ \\ \hline
$10,000$ & $0.99$ & $0.98$ & $0.2$ & $0.1$ & $16$ & $93.36\%$ & $73.88\%$ \\
\hline
\end{tabular}
}
\end{table}

Figure~\ref{fig:mnist-epoch-rp} shows the GMIA performance on models trained with 1k epochs and 10k epochs respectively. Reducing the training epoch did not eliminate membership privacy risk because a few records in the dataset were still identified with high precision. Specifically, when we increased the neighborhood threshold $\delta$ from $0.2$ to $0.3$, the $4$ vulnerable target records were identified with a precision of $1$ and a recall of $2.5\%$. 

Moreover, as we increased the maximum training epoch, the model's testing accuracy increased, indicating an improvement in model generalization and model utility. However, this small improvement in model utility came at a huge cost for privacy---it increased the attack precision from $72.27\%$ to $93.36\%$ and recall from $6.14\%$ to $73.88\%$.

\subsection{Influence of Regularization}
\label{subsec:l2reg}
Regularization is a common method for improving model generalization. It is shown to be an effective defense against the prior MIA~\cite{shokri2016membership}. To study its effectiveness on GMIA, we applied L$2$ regularization on neural networks trained on MNIST set even though the models were \emph{not overfitted}. In doing so, we limited the model capacity which increased the risk of underfitting. Specifically, when the regularization coefficient $\lambda$ went from $0.001$ to $0.01$, testing accuracy decreased by $0.01$ indicating that the model might be underfitted due to over regularization.

Table~\ref{tab:l2} and Figure~\ref{fig:mnist-l2} shows the model accuracy and GMIA performance before and after applying L2 regularization with varying coefficients $\lambda$. Applying regularization reduced the number of vulnerable target records in the dataset, but did not completely eliminate the privacy risk. The remaining vulnerable records were attacked with high precision. Specifically, when L2 regularization was applied with coefficient $\lambda = 0.01$, we still identified $1$ vulnerable target record, which was inferred with precision close to $1$ and a recall of $4\%$. 

\begin{table}[h]
{\scriptsize
\vspace{5pt}
\centering
\caption{\fontsize{9}{11}\selectfont GMIA w.r.t. Regularization ($\delta =0.2$, $\beta = 2$, $p=0.01$)}
\label{tab:l2}
\begin{tabular}{C{2cm}C{1.1cm}C{0.6cm}C{1.1cm}C{1.1cm}C{1.1cm}}
\hline
Regularization Coefficient $\lambda$ & Training Acc. & Test Acc. & \# of Targets & prec. & recall \\
\hline
$0$ & $0.99$ & $0.98$ & $52$ & $90.84\%$ & $68.31\%$ \\ \hline
$0.001$ & $0.99$ & $0.99$ & $1$ & $1$ & $54.8\%$ \\ \hline
$0.01$ & $0.98$ & $0.98$ & $1$ & $93.36\%$ & $4\%$ \\
\hline
\end{tabular}
}
\end{table}

Like reducing training epoch, applying regularization mitigated the model's privacy risk but did not eliminate the risk. Moreover, since the most vulnerable record was identified with high precision, regularization may not be a good approach when the data owner wants to provide privacy protection for \emph{all} individuals whose records are in the dataset. 

\section{Discussion}


\subsection{Understanding GMIA}

\noindent\textbf{Intuitions}. As mentioned in Section~\ref{sec:understanding}, MIA can succeed by querying a record $q$ if the target record has a unique influence on the predictions on $q$. Specifically, when we attack by querying the target record, the target record $r$ is vulnerable to MIA when there is a non-overlapping area between the two distributions: the distribution of predictions on $r$ when $r$ is not used to train the model and the distribution of predictions on $r$ when $r$ is used. To verify our understanding, we plot the distribution of predictions on a vulnerable record $r^*$ (Figure~\ref{fig:vulneralbe_record_mnist} in the Appendix) in the MNIST dataset. Firgure~\ref{fig:distribution} in the Appendix shows this distribution. In section~\ref{sec:eval}, the membership of $r^*$ is inferred with a precision of $1$ and a recall of $1$ when directly querying $r^*$. This high vulnerability is explained by the fact that there is almost no overlapping between the distributions of predictions on $r$ when $r$ is included and not included in the training dataset.

\vspace{2pt}\noindent\textbf{Limitations}. In the meantime, our current design of GMIA is preliminary. Our techniques for identifying outliers cannot find all vulnerable instances: it is possible that some instances not considered to be outliers by our current design still exert unique influences on the model, which need to be better understood in the follow-up research. Moreover, the current way to search for the for the enhancing records, through filtering out random queries, is inefficient, and often does not produce any results. More effective solutions could utilize a targeted search based upon a better understanding about the relations between the target record and other records. Also in line with the prior research~\cite{shokri2016membership}, we assume the adversary to either know the training algorithm or have black-box access to the training algorithm as an oracle. In practice, we may not be able to use the model identical to the target to train our references.  Our preliminary study shows that it is still possible to attack some vulnerable instances in an online target model (though at lower success rate) using off-line models. How to make this more effective needs further investigation. Fundamentally, it remains unclear how much information about the training set is leaked out through querying a machine learning model and whether more sensitive techniques can be developed to capture even a small signals for a record's unique impact. 

\subsection{Mitigation}


\noindent\textbf{Generalization and perturbation}. As mentioned earlier, generalization has limited effect on mitigating GMIA: as demonstrated in our study, even after applying the L2 regularization (with a coefficient of $0.01$), still a vulnerable record in MNIST dataset can be attacked with a precision of $1$ (Section~\ref{subsec:l2reg}). In the meantime, adding noise to the training set or to the model to achieve differential privacy can suppress the information leak~\cite{dwork2006differential}. However, in the presence of high-dimensional data, which is particularly vulnerable to our attack, perturbation significantly undermines the utility of the model before its privacy risk can be effectively controlled~\cite{johnson2013privacy}. As an example, a recent study reports that a differentially-private stochastic gradient descent (SGD) only has an accuracy of $0.6$ with $\epsilon=1$ and an accuracy of $0.5$ with $\epsilon=0.5$~\cite{mcmahan2017learning} on the MNIST dataset. So we believe that a practical solution should apply generalization and perturbation together with proper training set selection, detecting and removing those vulnerable training instances. 

\vspace{2pt}\noindent\textbf{Training record selection}. We believe that there is a fundamental contention between selecting useful training instances, which bring in additional information, and suppressing their unique influence to protect their privacy. An important step we could take here is to automatically identify outliers and drop those not contributing much to the utility of the model. To this end, new techniques need to be developed to balance the risk mitigation and the utility reduction for those risky instances. A machine learning model could be built to automatically decide whether an instance should be in the training set or not. 
\section{Related Work}
\noindent\textbf{Attacks on Machine Learning Models}. Different attacks against machine learning models have been proposed in recent years. For example, reverse engineering attacks~\cite{tramer2016stealing,hitaj2017deep} steal model parameters and structures; adversarial learning~\cite{evtimov2017robust,szegedy2013intriguing,liu2016delving,kos2017adversarial} generates misleading examples that will be misclassified by the model; model inversion attacks~\cite{fredrikson2014privacy,cormode2011personal} infer the features of a record based on the model's predictions on it; membership inference attacks~\cite{shokri2016membership} infer the presence of a record in the model's training dataset.


\noindent\textbf{Privacy and Model Generalization}. 
There is a connection between privacy and model generalization. Differential privacy can improve model generalization when data is reused for validation~\cite{dwork2006differential}. Moreover, the prior membership inference attack~\cite{shokri2016membership} achieves high accuracy on highly overfitted models while barely works on non-overfitted ones. Previous research also points out that privacy leakage can happen on non-overfitted models \emph{when the adversary has control over the training algorithm}. Specifically, the adversary can encode private information of the training dataset into the predictions of well-generalized models~\cite{song2017machine}. These two attacks~\cite{shokri2016membership,song2017machine} can be formalized under a uniform theoretical framework~\cite{yeom2017unintended}. The risk of membership inferences can be empirically measured based on the influence of each training record~\cite{long2017towards}.

\noindent\textbf{Privacy-Preserving Machine Learning} Differential privacy~\cite{dwork2006differential} is a prominent way to formalize privacy against membership inference. It has been applied to various machine learning models including decision trees~\cite{jagannathan2009practical}, logistic regression~\cite{zhang2012functional}, and neural networks~\cite{abadi2016deep,shokri2015privacy}. However, there are no generic methods to achieve differential privacy for all useful machine learning models. More importantly, even if these methods are developed, their applications to real-world machine learning problems may significantly decrease the accuracy of the models, and thus will reduce their utility~\cite{alvim2011differential}. 


\section{Conclusion}

In this paper, we take a step forward on understanding the information leaks from machine learning models. Our study demonstrates that overfitting contributes to the information leaks but is \textit{not} the fundamental cause of the problem. This understanding is achieved through a series of membership inference attacks on well-generalized models, discovering vulnerable instances (cancer patients, images, and individual data) even without directly querying the vulnerable target records, and even in the presence of regularization protection.  Our study highlights the contention between selecting informative training instances and preventing their identification through their unique influences on the model, and points to the direction of using training data analysis and selection to complement existing approaches. 



{\footnotesize \bibliographystyle{acm}
\bibliography{refs}}
\vfill

\pagebreak
\section*{Appendix}
\vspace{3pt}\noindent\textbf{Training and Testing Accuracy of Target Models}.
Table~\ref{tab:acc} shows the training and testing accuracy of the target models in our attacks. All the target models are well-generalized models with difference between training and testing accuracy smaller than $0.1$.
\begin{table}[h]
{\scriptsize
\centering
\caption{\fontsize{9}{11}\selectfont Training and Testing Accuracy of Target Models}
\label{tab:acc}
\begin{tabular}{C{2.5cm}C{2.5cm}C{2.5cm}}
\hline
Dataset (Model) & Training Accuracy & Test Accuracy \\
\hline
Adult & $0.85 \pm 0.01$  & $0.85$ \\ \hline
Cancer & $0.95 \pm 0.04$ & $0.94 \pm 0.03$ \\ \hline
MNIST & $0.99$ & $0.98$ \\ \hline
Adult(Google) & $0.84 \pm 0.03$ & $0.84 \pm 0.02$ \\ \hline
MNIST(Google) & $0.90$ & $0.90$ \\
\hline
\end{tabular}
}
\end{table}

\begin{figure}[h]
    \centering
\includegraphics[width=0.3\textwidth]{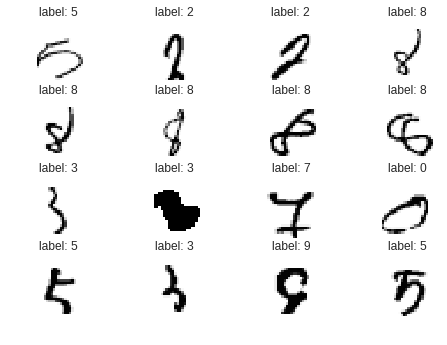}
    \caption{Vulnerable Examples in MNIST Dataset}\label{fig:mnist-outliers}
\end{figure}

\vspace{3pt}\noindent\textbf{Vulnerable Records in MNIST Dataset}.
To study what kinds of records are vulnerable to GMIA, we plotted the vulnerable target records selected from MNIST dataset with $\delta=0.2$ and $\beta=0.1$ (Figure~\ref{fig:mnist-outliers}). As we expected, some of the vulnerable target records are outliers in the dataset. However, some vulnerable examples actually increase model utility by providing rare but useful features for the classification task. For example, the images of digit $8$ written in different directions may help a model on recognizing similar written digits in testing examples. However, since these images are rare in the dataset, they have a unique influence on the target models, making them vulnerable to GMIA, and the fact that this influence is useful in predicting unseen examples does not mitigate the risk. 

\begin{figure}[h]
    \centering
	\includegraphics[width=0.3\textwidth]{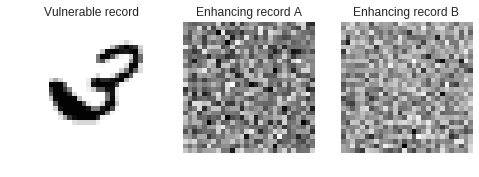}   
    \caption{Vulnerable record from MNIST with its two enhancing records.}\label{fig:vulneralbe_record_mnist}
\end{figure}

\vspace{3pt}\noindent\textbf{Indirect Inference on MNIST Dataset}. To study the correlation between a target record and its enhancing records in indirect inferences, we plotted the target record with its enhancing records in the MNIST dataset~\ref{fig:vulneralbe_record_mnist}. Surprisingly, the enhancing records seem like images of random noise and by no means represent the target record. 

\begin{figure}[h]
    \centering
	\includegraphics[width=0.3\textwidth]{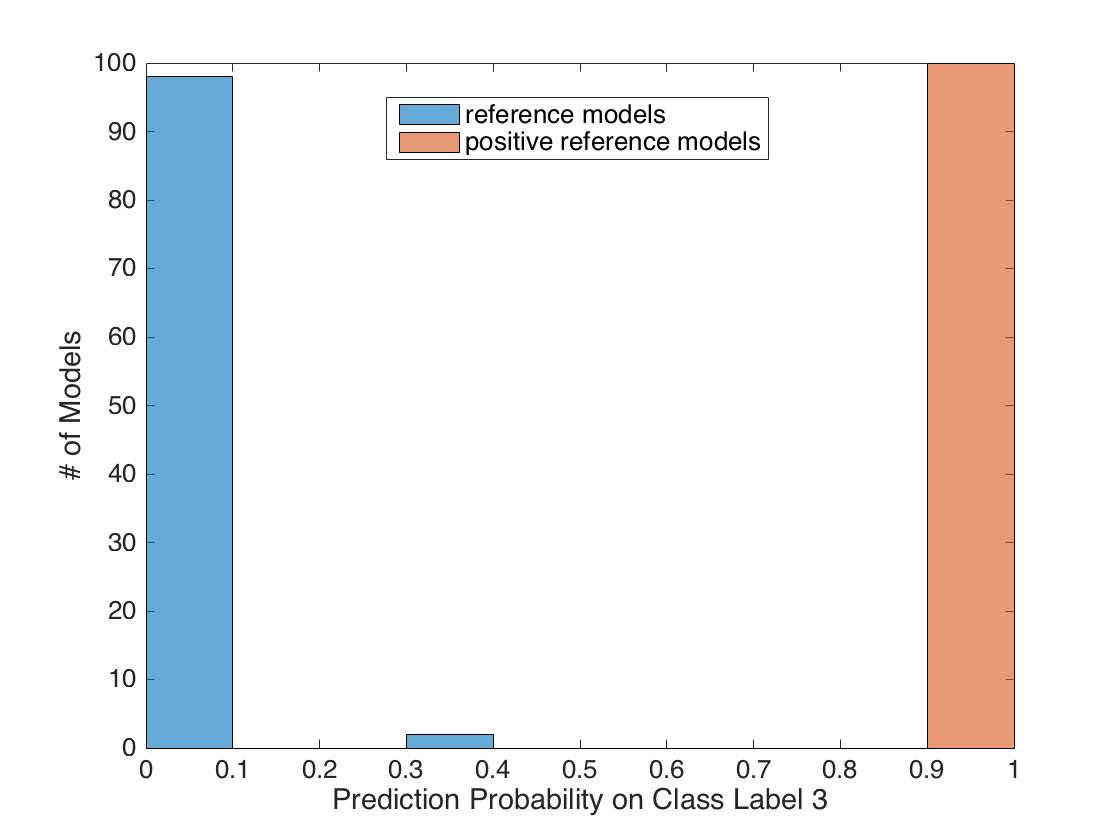}   
    \caption{The histogram of predictions on $r^*$ when $r^*$ is in the training dataset (i.e., positive reference models) v.s. not in the training dataset (i.e., reference models)}\label{fig:distribution}
\end{figure}

\vspace{3pt}\noindent\textbf{Intuitions on GMIA}. Figure~\ref{fig:distribution} shows a vulnerable record's influence on the machine learning model's predictions on itself. The image of the record is plotted in Figure~\ref{fig:vulneralbe_record_mnist}. All the positive reference models (i.e., reference models trained with the target record) predict high probability for the correct class label while all the reference models (i.e., models trained without the target record) predict low probability for the correct class label. This difference allows us to successfully infer the presence of this record in the training dataset.

\vspace{3pt}\noindent\textbf{Comparison with the Prior MIA}.
To compare with the prior MIA~\cite{shokri2016membership}, we reproduced the attack in~\cite{shokri2016membership} on the same target models and same vulnerable records in GMIA. Specifically, we trained one attack classifier per class for each dataset. The attack classifiers are neural networks with one hidden layer of 64 units. We used \texttt{ReLU} as the activation function and \texttt{SoftMax} as the output layer. 
We only performed the attack when the probability given by the attack classifier was higher than a certain threshold (called attack confidence threshold). We evaluated the performance of the attack under various attack threshold as shown in Table~\ref{tab:prior_attack}. The attack precision was relatively low (e.g. $<70\%$) on all three datasets even when a high attack confidence threshold was used.
\begin{table}[h]
{\scriptsize
\centering
\vspace{5pt}
\caption{\fontsize{9}{11}\selectfont Performance of the Prior MIA on the Same Target Models}
\label{tab:prior_attack}
\begin{tabular}{C{1.75cm}C{2cm}C{1.75cm}C{1.5cm}}
\hline
Dataset & Attack Confidence Threshold & Attack Precision & Attack Recall \\
\hline
\multirow{2}{*}{\specialcell{Cancer \\ ($3$ records)}} 
& $0.8$ & $50.25\%$ & $40\%$ \\ 
& $0.9$ & - & $0$ \\ 
\hline
\multirow{2}{*}{\specialcell{Adult \\ ($13$ records)}} 
& $0.6$ & $66.67\%$ & $4.92\%$ \\ 
& $0.7$ & - & $0$ \\
\hline
\multirow{3}{*}{\specialcell{MNIST \\ ($16$ records)}} & $0.6$ & $50\%$ & $56.25\%$ \\
& $0.7$ & $19.6\%$ & $6.25\%$ \\
& $0.8$ & - & $0$ \\
\hline
\end{tabular}
}
\end{table}

\end{document}